\documentclass[aps,amsmath,amssymb,amsfonts,12pt]{revtex4}
\usepackage{graphicx}
\usepackage{wasysym}
\newcommand{\bm}{\boldmath}
\newcommand{\D}{\displaystyle}
\newcommand{\vek}[1]{\mbox{\bm ${#1}$}}
\begin{document}
\thispagestyle{empty}
\renewcommand{\theequation}{\arabic{section}.\arabic{equation}}
\vspace{3cm}
\begin{center}{\large\bf The high-density electron gas: How its momentum distribution $n(k)$ and its static structure factor $S(q)$ 
are mutually related through the off-shell self-energy $\Sigma (k,\omega)$}\\
\end{center}
\begin{center}{\sc P. Ziesche} \\
{Max-Planck-Institut f\"ur Physik komplexer Systeme \\
N\"othnitzer Str. 38, D-01187 Dresden, Germany}\\
\end{center}
\begin{center}
PACS 71.10.Ca, 05.30.Fk
\end{center}
\date{draft of \today}
\vspace{2cm}
\begin{abstract}
\noindent
It is shown {\it in detail} $how$ the ground-state self-energy $\Sigma(k,\omega)$ of the spin-unpolarized uniform electron 
gas (with the density parameter $r_s$) in its high-density limit $r_s\to 0 $ determines: the momentum distribution 
$n(k)$ through the Migdal formula, the kinetic energy $t$ from $n(k)$, the potential energy $v$ through the Galitskii-Migdal formula, 
the static structure factor $S(q)$ from $e=t+v$ by means of a Hellmann-Feynman functional derivative. The ring-diagram partial 
summation or random-phase approximation is extensively used and the results of Macke, Gell-Mann/Brueckner, Daniel/Vosko, Kulik, and 
Kimball are summarized in a coherent manner. There several identities were brought to the light.  
\end{abstract}
\maketitle

\newpage

\section{Introduction}
\setcounter{equation}{0}
\noindent
Although not present in the Periodic Table the homogeneous electron gas (HEG) is still an important and so far unsolved 
model system for electronic structure theory, cf. e.g. \cite{Tos}. In its spin-unpolarized version, the 
HEG ground state is characterized by only one parameter $r_s$, such that a sphere with the radius 
$r_s$ contains {\it on average} one electron \cite{Zie1}. It determines the Fermi wave number as 
$k_{\rm F}=1/(\alpha r_s)$ in atomic units (a.u.) with $\alpha =[4/(9\pi)]^{1/3}\approx 0.521062$ and it measures simultaneously 
both the interaction strength and the density such that high density corresponds to weak interaction and 
hence weak correlation \cite{foo}. For recent papers on this limit cf. \cite{Cio0,Zie2,Zie3,Mui,Cio,Zie4,Gla2,Zie5}. 
Usually the 
total ground-state energy per particle is written as (here and in the following are wave numbers measured in units of 
$k_{\rm F}$ and energies in units of $k_{\rm F}^2$)
 \begin{equation}\label{a1}
e=e_0+e_{\rm x}+e_{\rm c}, \quad e_0=\frac{3}{5}\cdot \frac{1}{2}, \quad 
e_{\rm x}=-\frac{3}{4}\cdot\frac{\alpha r_s}{\pi}, \quad e_{\rm c}=(\alpha r_s)^2[a\ln r_s +(b+b_{2{\rm x}})+O(r_s)],
\end{equation}  
where $e_0$ is the energy of the ideal Fermi gas, $e_{\rm x}$ is the exchange energy in lowest (1st) order (the corresponding direct
term is zero, because the system is neutral), and $e_{\rm c}$ is referred to as correlation energy. The constants $a$ and $b$ arise 
from the ring-diagram summation as explained in the following. Naively one should 
expect that in the high-density limit the Coulomb repulsion $\epsilon^2/r$ \cite{foo}
can be treated as perturbation. But in the early theory of the HEG, Heisenberg has shown \cite{Hei}, that ordinary
perturbation theory with $e_{\rm c}=e_2+e_3+\cdots$ and $e_n\sim (\alpha r_s)^n$, where the subscript $n$ is the perturbation order, 
does not apply. Namely, in 2nd order,
there is a direct term $e_{2{\rm d}}$ and an exchange term $e_{2{\rm x}}$, so that $e_2=e_{2{\rm d}}+e_{2{\rm x}}$. Unfortunately
the direct term $e_{2{\rm d}}$ logarithmically diverges along the Fermi surface (i.e. for
vanishing transition momenta $q$): $e_{2{\rm d}}\to\ln q$ for $q\to 0$ \cite{Hei}. This failure of perturbation theory has been
repaired by Macke \cite{Ma} with an appropriate partial summation of higher-order terms $e_{3{\rm r}},e_{4{\rm r}},
\cdots$ (the subscript ``r'' means ``ring diagram'') up to infinite order. The result is Eq. (\ref{a1}) with 
$a=(1-\ln 2)/\pi^2\approx 0.031091$ after Macke \cite{Ma} and $b\approx -0.0711$ after Gell-Mann and Brueckner 
\cite{GB}. The latter means, that there is a pure 2nd-order remainder of the ring-diagram summation in addition to the non-analyticity
$r_s^2\ln r_s$. A consistent description up to this order requires to take into account all other terms of the same order, i.e. 
$e_{\rm 2x}\sim r_s^2$. After Onsager, Mittag, and Stephen \cite{Ons} it is $e_{\rm 2x}=(\alpha r_s)^2b_{\rm 2x}$ with       
$b_{2{\rm x}}=(1/6)\ln 2-3 \zeta(3)/(2\pi)^2\approx +0.02418$. Thus part of the direct term $b$ is compensated by the exchange term 
$b_{\rm 2x}=-0.34\ b\ \curvearrowright\ b+b_{\rm 2x}=0.66\ b$. The physically plausible partial summation of
higher-order perturbation terms used by Macke, respectively Gell-Mann and Brueckner is called ring-diagram summation with its 
characteristic particle-hole-pair excitations ($\vek k\to\vek k+\vek q, |\vek k|<1,|\vek k+\vek q|>1$, $\vek q=$ momentum transfer),  
also known as the RPA = random phase approximation. It is the simplest approximation which simultaneously describes the closely related
phenomena of screening effects and plasma oscillations as well as plasmon propagation with dispersion. (For their damping one has
to go beyond RPA with local field corrections.) In the high-density limit, correlation ``c'' starts with RPA or ring-diagram terms 
``r'', symbolically written as c = r + $\cdots$. In the following the screening parameter $q_c^2=4\alpha r_s/\pi$ (being the
interaction strength $\alpha r_s$ times $4/\pi$) is used. In terms of this screening wave number $q_c$ the electron-gas plasma
frequency (measured in units of $k_{\rm F}^2$) is $\omega_{\rm pl}=q_c/\sqrt 3={\sqrt {4\alpha r_s/(3\pi)}}$.  The building elements
of the RPA Feynman diagrams are shown in Figs. 1a,b. The diagrams for $e_{\rm r}$ and $e_{\rm 2x}$ are in Figs. 1c,d, middle 
parts. \\ 

\noindent 
The non-analytical behavior of the total energy $e$ at the high-density limit carries over to its kinetic and potential components, 
$t$ respectively $v$, through the virial theorem \cite{Mar} and to the chemical potential $\mu$ through the Seitz theorem \cite{Sei}: 
\begin{equation}\label{a2}
v=r_s\frac{d}{d r_s}e\ ,\quad  t=-r_s^2\frac{d}{dr_s}\frac{1}{r_s}e\ , \quad \mu=\left(\frac{5}{3}-\frac{1}{3}r_s\frac{d}{dr_s}\right)e\ .
\end{equation}
In the high-density limit, this means
\begin{equation}\label{a3}
t=t_0+t_{\rm x}+t_{\rm c}, \quad t_0=\frac{3}{5}\cdot \frac{1}{2}, \quad t_{\rm x}=0, \quad
t_{\rm c}=-(\alpha r_s)^2[a \ln r_s+(a+b+b_{2\rm x})+O(r_s)]\ ,
\end{equation}
\begin{equation}\label{a4}
v=v_0+v_{\rm x}+v_{\rm c}, \quad v_0=0, \quad v_{\rm x}=-\frac{3}{4}\cdot\frac{\alpha r_s}{\pi}, \quad
v_{\rm c}=(\alpha r_s)^2[2a \ln r_s+(a+2b+2b_{2\rm x})+O(r_s)]\ ,
\end{equation}
\begin{equation}\label{a5}
\mu=\mu_0+\mu_{\rm x}+\mu_{\rm c}\ , \quad \mu_0=\frac{1}{2}\ , \quad \mu_{\rm x}=-\frac{\alpha r_s}{\pi}\ , \quad
\mu_{\rm c}=(\alpha r_s)^2\left[a \ln r_s+\left(-\frac{1}{3}a+b+b_{\rm 2x}\right)+O(r_s)\right]\ .
\end{equation}
Fundamental relations between the simplest quantum-kinematical quantities (momentum distribution $n(k)$ and static structure 
factor $S(q)$) and the energy components $t$ and $v$ are
\begin{eqnarray}\label{a6}
t&=&\int\limits_0^\infty d(k^3)\ n(k)\frac{k^2}{2}\ ,\quad \int\limits_0^\infty d(k^3)\ n(k)=1 \ ,\quad 0\leq n(k)\leq 1\ ,   \\
&& \quad z_{\rm F}=n(1^-)-n(1^+), \quad 0\leq z_{\rm F}=1-O(r_s)\ , \quad n(k\to \infty)\to\frac{A(r_s)}{k^8}+\cdots\ ,   \nonumber \\
v&=&-\frac{1}{3\cdot 4}\int\limits_0^\infty d(q^3)\ [1-S(q)]\frac{q_c^2}{q^2}\ ,\quad
S(q\to 0)=\frac{q^2}{2\omega_{\rm pl}}+\cdots\ ,\quad [1-S(q\to\infty)]\to\frac{B(r_s)}{q^4}+\cdots\ \nonumber \\ 
\end{eqnarray}
with the static 1-body quantity $n(k)$, the momentum distribution, and with the static 2-body quantity $S(q)$, the static structure 
factor 
(SSF). The Fourier transform of $1-S(q)$ is $1-g(r)$ with $g(r)\geq 0$ being the pair density (PD), see Sec.V. The SSF $S(q)$ behaves at 
transition momenta $|\vek q|=2$ non-analytically, because there occurs a topological change from two overlapping to two non-overlapping 
Fermi spheres. This causes asymptotic Friedel oscillations of the PD $g(r\to\infty)$, whereas cusp singularities $S(q\rightarrow 0)\sim q^{\rm odd}$ let emerge non-oscillatory asymptotic terms of $g(r\to \infty)$. 
For the asymptotic 
coefficients the sum rules 
\begin{equation}\label{a8}
A(r_s)=\frac{1}{2}\omega_{\rm pl}^4\ g(0) \quad {\rm and} \quad B(r_s)=2\omega_{\rm pl}^2\ g(0)\quad {\rm with}\quad 
1-g(0)=\frac{1}{2}\int\limits_0^\infty d(q^3)\ [1-S(q)]
\end{equation} 
hold with the on-top PD $g(0)=1/2-O(r_s)$ \cite{Kim2,YaKa} (for $n(k)$) and \cite{Kim1,Kim2,Ya} (for $S(q)$). $g(0)$ is given by 
the normalisation of $1-S(q)$ and describes short-range correlations together with the peculiar behavior of $g(r\ll 1/q_c)$, 
besides it determines the large-wave-number asymptotics of $n(k)$ and $S(q)$. \\

\noindent
In view of (\ref{a6}) and (1.7), one may ask, which peculiarities of these lowest-order quantum-kinematical quantities cause the 
non-analyticities of $t$ and $v$ \cite{Zie3}. The above mentioned 
drastic changes, when switching on the Coulomb interaction, show up in the redistribution of the non-interacting momentum distribution 
$n_0(k)=\Theta (1-k)$ within thin layers inside and outside the Fermi surface $|\vek k|=1$ and a remaining finite discontinuity 
$z_{\rm F}> 0$ (Migdal theorem \cite{Mig,Lu}). They show up also in the plasmon behavior of $S(q)$ within a small 
spherical region around the origin of the reciprocal space, what causes an inflexion point $q_{\rm infl}, S_{\rm infl}\sim 
\omega_{\rm pl}$. All these reconstructions describe the long-range correlation (screening, collective mode called plasmon), 
characteristic for the Coulomb interaction. They are treated in lowest order again by ring-diagram summations with the replacements 
$n_{\rm 2d}(k)\to n_{\rm r}(k)$ and $S_{\rm 1d}(q)\to S_{\rm r}(q)$. (Note, that $t_{\rm 2d}$ arises from $n_{\rm 2d}(k)$, but 
$v_{\rm 2d}$ from $S_{\rm 1d}(q)$.) How these replacements lead to the $r_s^2\ln r_s$ terms 
of $t$ and $v$ is shown in \cite{Zie3}. However, because the redistributions take place essentially only in the mentioned sensitive 
regions $||\vek k|-1|\ll q_c$ and $|\vek q|\ll q_c$ the r-terms approach the original d-terms far off these regions. 
Thus, to be consistent
up to this order, the corresponding x-terms must be taken into account: $n_{\rm c}(k)=n_{\rm r}(k)+n_{\rm 2x}(k)+\cdots$, 
$S_{\rm c}(q)=S_{\rm r}(q)+S_{\rm 1x}(q)+\cdots$. These x-terms compensate part of the r-terms, similarly 
as this is the case for
$e_{\rm c}$, $t_{\rm c}$, $v_{\rm c}$, $\mu_{\rm c}$ with the ratios of $b_{\rm 2x}$ to $b$, $a+b$, $a/2+b$, $-a/3+b$ being -0.34, -0.6,
-0.43, -0.3, respectively. The ring-diagram summation
for $n(k)$ has been developed in \cite{Da,Ku}, an analytical extrapolation is given in \cite{GGZ}, for the spin-polarized case see 
\cite{Zie8}. The ring-diagram summation for $S(q)$ 
has been done in \cite{Gli,Kim3}. In \cite{La} both $n(k)$ and $S(q)$ are considered on the same footing.  \\   

\noindent
Another relevant ground state property is Dyson's self-energy $\Sigma(k,\omega)$, a {\it dynamical} 1-body quantity. Its on-shell value 
(as a function of $r_s$) is related to the chemical-potential shift through the Hugenholtz-van Hove (Luttinger-Ward) theorem
$\mu-\mu_0= \Sigma(1,\mu)$ \cite{Hug}.   
Besides this, with the off-shell self-energy $\Sigma(k,\omega)$ also the 1-body Green's function
\begin{equation}\label{a9}
G(k,\omega)=\frac{1}{\omega-t(\vek k)-\Sigma(k,\omega)}\quad {\rm with}\quad  \Sigma(k,\omega)\to{\rm sign}(k-1)\ {\rm i}\delta\quad 
{\rm for}\quad r_s\to 0 \\
\end{equation} 
is known, from which follow the momentum distribution (Migdal formula \cite{Mig})
\begin{equation}\label{a10}
n(k)=\int\limits_{C_+}\frac{d\omega}{2\pi{\rm i}}\;{\rm e}^{{\rm i}\omega\delta}G(k,\omega)
\quad  \curvearrowright \quad \int\limits_0^\infty d(k^3)\ n(k)=1, \quad 
t=\int\limits_0^\infty d(k^3)\ n(k)\ \frac{k^2}{2} 
\end{equation}
and the potential energy (Galitskii-Migdal formula \cite{Gal}, for its use in total-energy calculations cf. \cite{Miy} and refs. therein)
\begin{equation}\label{a11}
v=\frac{1}{2}\int\limits_0^\infty d(k^3)\int\limits_{C_+}\frac{d\omega}{2\pi{\rm i}}\;{\rm e}^{{\rm i}\omega\delta}G(k,\omega)
\Sigma(k,\omega)   
\end{equation}
with $\delta{_> \atop ^{\to}}0$ and  $C_+$ means the closing of the contour in the upper complex $\omega$-plane.  \\ 

\noindent
Note that $\Sigma(k,\omega)$ and thus also $G(k,\omega)$, $n(k)$ and $t$, as well as $v$ are functionals of $t(\vek k)$ and $v(\vek q)$. 
Supposed the ground state energy $e$ is available as such a {\it functional}, then the (generalized)
Hellmann-Feynman theorems \cite{Mar} 
\begin{equation}\label{a12}
n(k)=\frac{4\pi}{3}\frac{\delta e}{\delta t(\vek k)}\ , \quad \quad S(q)-1=16\pi\frac{\delta e}{\delta v(\vek q)}
\end{equation}
hold, cf. App. A. These are equivalent writings of expressions given in \cite{Da,La} for $n(k)$ and in 
\cite{Ya,Kim3} for $S(q)$. The quantities 
$e$, $n(k)$, $t$, $S(q)$, and $v$ result as functions of $r_s$ from the replacements $t(\vek k)\to k^2/2$ and
$v(\vek q)\to q_c^2/q^2$. These fundamental relations 
permit the following procedure (see Fig. 2): If $\Sigma(k,\omega)$ is available as a functional 
of $t(\vek k)$ and $v(\vek q)$ from perturbation theory or otherwise, then $n(k)$ can be calculated with the Migdal formula 
(\ref{a10}). Therefrom follows $t$ with (\ref{a6}) and $v$ with the Galitskii-Migdal formula
(\ref{a11}). Finally from their sum $t+v=e$ and the functional derivative (\ref{a12}) the SSF $S(q)$  results (and $n(k)$ may be 
checked once more for consistency). So the dynamical 1-body quantity $\Sigma(k,\omega)$ provides the static 2-body quantity $S(q)$. \\

\noindent
Whereas in \cite{Zie5} the {\it on-shell} self-energy $\Sigma (1,\mu)$ has been studied, here the {\it off-shell} self-energy 
$\Sigma(k,\omega)$ is considered. The problem in \cite{Zie5} was, to find out the correct diagrammatic sum for $\Sigma(1,\mu)$ on the rhs
of the Luttinger-Ward theorem, which makes it an identity in the high-density limit $r_s\to 0$. The answer: the $GW$ approximation with 
$W$= RPA and $G$
= an appropriately renormalized particle-hole line yield the correct $r_s$-behavior, which agrees with the lhs as it follows from 
$\mu(r_s\to 0)$, cf. (\ref{a5}). Here it is shown in detail how the off-shell self-energy $\Sigma(k,\omega)$ for the model case $r_s\to 0$ 
yields $n(k)$, $t$, $v$, and $S(q)$ step by step according to the procedure of Fig. 2. So it is shown how $n(k)$ and $S(q)$, which have
their common origin in the 2-body density matrix, are indirectly linked mutually through the self-energy $\Sigma(k,\omega)$. \\ 

\noindent
Perturbation theory for $\Sigma(k,\omega)$ means $\Sigma=\Sigma_{\rm x}+\Sigma_{\rm c}$ with $\Sigma_{\rm c}=\Sigma_2+\Sigma_3+\cdots$, 
where $\Sigma_2=\Sigma_{\rm 2d}+\Sigma_{\rm 2x}$. The divergence of $\Sigma_{\rm 2d}$ is corrected by the ring-diagram summation (RPA):
$\Sigma_{2d}+\cdots\to\Sigma_{\rm r}$, so $\Sigma _{\rm c}=\Sigma_{\rm r}+\Sigma_{\rm 2x}+\cdots$. The building elements of the RPA 
Feynman diagrams (cf. Figs. 1a and 1b) are the Coulomb repulsion 
$\pi^2v(\vek q)\to\pi^2q_c^2/q^2$ with the coupling constant $q_c^2=4\alpha r_s/\pi$ and the 
one-body Green's function of free electrons with $t(\vek k)\to k^2/2$,
\begin{equation}\label{a13}
G_0(k,\omega)=\frac{\Theta(k-1)}{\omega-t(\vek k)+{\rm i}\delta}+
\frac{\Theta(1-k)}{\omega-t(\vek k)-{\rm i}\delta}\ , \quad  {\mbox{ $\delta{_> \atop ^{\to}}0$}}\ .
\end{equation}
From $G_0(k,\omega)$ follows the particle-hole propagator $Q(q,\eta)$ in RPA according to
\begin{equation}\label{a14}
Q(q,\eta)=-\int\frac{d^3k}{4\pi}\int\frac{d\omega}{2\pi{\rm i}}G_0(k,\omega)
G_0(|{\mbox{\bm $k$}}+{\mbox{\bm $q$}}|,\omega+\eta)
\end{equation}
with the result
\begin{equation}\label{a15}
Q(q,\eta)=\int\limits_{|\vek k|<1,\ |\vek k+\vek q|>1} \frac{d^3k}{4 \pi} \left[\frac{1}{t(\vek k+\vek q)-t(\vek k) -\eta-{\rm i}\delta}+
\frac{1}{t(\vek k+\vek q)-t(\vek k)+\eta-{\rm i}\delta}\right]\ .
\end{equation}
The denominators contain the excitation energy to create a hole with $\vek k$ inside the Fermi sphere and a particle with $\vek k+\vek q$
outside the Fermi sphere. Thus, $Q(q,\eta)$ is a functional of $t(\vek k)$ with the functional derivative (\ref{A5}) and with 
$R(q,u)=Q(q,{\rm i}qu)$ defining a real function, see (\ref{B1}).
In the following the self-energy contributions $\Sigma_{\rm x}$ (Sec. II), $\Sigma_{\rm r}$ (Sec. III), 
$\Sigma_{2{\rm x}}$ (Sec.VI) are explicitly given as they result from the diagram rules and it is derived, what follows from them:
$n(k),\ t,\ v,\ e,\ S(q)$ . For (here not considered) issues as $S(q,\omega)$, $\varepsilon(k,\omega)$, $GW$-approximation etc. cf. e.g.
\cite{Nech,Miy,Hol,Ara,Bro} and refs. therein. Sec. V deals with the short-range correlation following from $S(q)$, cf. (\ref{a8}). 
Although the 
high-density spin-unpolarized HEG is only a marginal corner in the complex field of electron correlation, its ring-diagram summation
 gives deeper insight through rigorous theorems, how energies and low-order quantum-kinematics are functionally related, what
should be of a more general interest.  

\section{First order}
\setcounter{equation}{0}

\noindent
The 1st-order direct term vanishes
because of the neutralizing positive background. So the expansion of $\Sigma(k,\omega)$ starts with the 1st-order exchange term 
\begin{equation}\label{b1}
\Sigma_{\rm x}(k,[v(\vek q)])=-\int\limits_{|\vek k+\vek q|<1} \frac{d^3q}{(2\pi)^3}\ \pi^2 v(\vek q)\ =
-\frac{1}{8\pi}\int\limits_{|\vek k+\vek q|<1} d^3q\ v(\vek q).
\end{equation}
Its peculiarity is, that it does not depend on $\omega$, what makes $n_{\rm x}(k)$ vanishing identically, because of
Eq. (\ref{a10}) with $G\approx G_0\Sigma_{\rm x}G_0$. This is in agreement with $t_{\rm x}=0$ as a consequence of the 
virial theorem (\ref{a2}). One may therefore conclude, that all the features of $n(k)$ start with the 2nd order. But this is not true.
The peculiar redistribution of the non-interacting momentum distribution $n_0(k)=\Theta(1-k)$ due to the Coulomb repulsion 
makes the dicontinuity jump $z_{\rm F}=n(1^{-})-n(1^+)$ to deviate from its non-interacting value of 1 in 1st order: 
$z_{\rm F}=1-0.18\ r_s+\cdots$. But the corresponding kinetic energy starts with $t_{\rm c}\sim r_s^2\ln r_s$. -  The potential energy 
$v_{\rm x}$ follows again as a functional of $v(\vek q)$ from the Galitskii-Migdal formula (\ref{a11}):
\begin{equation}\label{b2}
v_{\rm x}[v(\vek q)]= \frac{3}{8\pi}\int\limits_{|\vek k|<1}d^3k\ \Sigma_{\rm x}(k,[v(\vek q)])= 
-\frac{3}{(8\pi)^2}\int\limits_{|\vek k|<1, |\vek k+\vek q|<1}d^3k\int d^3q\ v(\vek q)\ . 
\end{equation}
This is, because of $t_{\rm x}=0$, also the total energy in this order: $e_{\rm x}[v(\vek q)]=v_{\rm x}[v(\vek q)]$. 
Thus the functional derivative (\ref{a12}) yields
\begin{equation}\label{b3}
S_0(q)=1-\frac{3}{4\pi} \int\limits_{|\vek k|<1,|\vek k+\vek q|<1}d^3k= 
\left[\frac{3}{2}\frac{q}{2}-\frac{1}{2} \left(\frac{q}{2}\right)^3\right]\Theta(2-q)+\Theta(q-2).
\end{equation}
The integral arises (for $q\leq 2$) from two overlapping Fermi spheres. For $q\geq 2$ they do not overlap, thus a 
drastical change of the topology occurs, when passing $q=2$. After Fourier transformation, see (\ref{e1}), this non-analyticity causes 
the asymptotic Friedel oscillations of the non-interacting PD $g_0(r\to \infty)-1\sim \cos 2r, \sin 2r$. 
Another non-analyticity, namely the cusp singularities $S_0(q\to 0)\sim q, q^3$ make the non-oscillatory terms of 
$g_0(r\rightarrow\infty)$.
The above integral $\int_{\cdots}d^3k$ is just the volume of two calottes 
with the height $h=1-q/2$, hence $\int_{\cdots} d^3k = 2\cdot\frac{\pi}{3}h^2(3-h)$.
If in Eqs. (\ref{b1}) and (\ref{b2}) the general interaction line $v(\vek q)$ is replaced by the Coulomb interaction $q_c^2/q^2$, then
\begin{equation}\label{b4}
\Sigma_{\rm x}\left(k,\left[\frac{q_c^2}{q^2}\right]\right)=-\frac{q_c^2}{4}\ \left(1+\frac{1-k^2}{2k} \ln\left|\frac{k+1}{k-1}\right|\right) \quad 
{\rm and} \quad v_{\rm x}=-\frac{3}{16}q_c^2 
\end{equation}
turn out. Besides $\Sigma_{\rm x}(1)=-q_c^2/4=(4/3)e_{\rm x}$. Note $e_{\rm x}=v_{\rm x}$ in agreement with the virial theorem 
(\ref{a2}). This shows how the procedure of Fig. 2 works in lowest order.

\section{Second Order: The direct term $\Sigma_{\rm 2d}$ and its RPA correction}
\setcounter{equation}{0}

\subsection{The self-energy $\Sigma_{\rm r}(k,\omega)$}

\noindent
In 2nd order there is a direct term (d) and an exchange term (x), see Figs. 1c,d, left: 
$\Sigma_2(k,\omega)=\Sigma_{2{\rm d}}(k,\omega)+\Sigma_{2{\rm x}}(k,\omega)$.
The direct term diverges along the Fermi surface, i.e. for vanishing transition momenta $q_0\to 0$.
This flaw is repaired by the ring-diagram summation (Fig. 1c, left) with the result  
\begin{equation}
\Sigma_{{\rm r}}(k,\omega)=\frac{1}{8\pi}\int\limits_{q>q_0}d^3q \int\frac{d\eta}{2\pi
{\rm i}} \; \frac{v^2(\vek q)Q(q,\eta)}{1+v(\vek q)Q(q,\eta)}\ G_0(|{\mbox{\bm $k$}}+{\mbox{\bm $q$}}|,\omega+\eta), \;
{\mbox{ $q_0{_> \atop ^{\to}}0$}}\ .  \nonumber \\
\end{equation}
Note, that this defines a functional of $v(\vek q)$ and note also   
\begin{equation}
\frac{v(\vek q)}{1+v(\vek q)Q(q,\eta)}\to\frac{q_c^2}{q^2+q_c^2Q(q,\eta)} \quad {\rm for} \quad v(\vek q)\to\frac{q_c^2}{q^2}\ ,
 \nonumber 
\end{equation}
where the screening (or ``Yukawa'') term $q_c^2Q(q,\eta)$ in the denominator makes the bare Coulomb 
repulsion renormalized and removes the above mentioned divergence of $\Sigma_{\rm 2d}(k,\omega)$. $G_0(k,\omega)$ and $Q(q,\eta)$ are
functionals of $t(\vek k)$, see (\ref{a13}), (\ref{a15}). Use of Eq. (\ref{a13}) leads to 
\begin{eqnarray}\label{c1} 
\Sigma_{\rm r}(k,\omega)&=&
\frac{1}{8\pi}\int d^3q\int \frac{d\eta}{2\pi{\rm i}}\
\frac{v^2(\vek q)Q(q,\eta)}{1+v(\vek q)Q(q,\eta)}\times \nonumber \\
&\times&\left[\D\frac{\Theta(|{\vek k}+{\mbox{\bm $q$}}|-1)}
{\omega+\eta-\frac{1}{2}k^2-{\mbox{\bm $q$}}\cdot(\vek k+\frac{1}{2}\mbox{\bm $q$})+{\rm i}\delta}+
\D \frac{\Theta(1-|{\vek k}+{\mbox{\bm $q$}}|)}{\omega+\eta-\frac{1}{2}k^2-{\mbox{\bm $q$}}\cdot(\vek k+\frac{1}{2}
\mbox{\bm $q$})-{\rm i}\delta}\right] .
\end{eqnarray} 
In the following it is shown, how $n_{\rm r}(k)$ and $t_{\rm r}$ result from this $\Sigma_{\rm r}(k,\omega)$.

\subsection{How $n_{\rm r}(k)$ results from $\Sigma_{\rm r}(k,\omega)$ and $t_{\rm r}$ from $n_{\rm r}(k)$}

\noindent
According to the Migdal formula (\ref{a10}), $n(k)$ follows from $G(k,\omega)$. In RPA it is
\begin{equation}\label{c2} 
G(k,\omega)=G_0(k,\omega)+G_0(k,\omega)\Sigma_{\rm r}(k,\omega)G_0(k,\omega)+\cdots
\end{equation}
The first term yields the momentum distribution of the ideal Fermi gas
\begin{equation}\label{c3}
n_0(k)=\int\limits_{C_+}\frac{d\omega}{2\pi{\rm i}}\left[\frac{\Theta(k-1)}{\omega-\frac{1}{2}k^2+{\rm i}\delta}+
\frac{\Theta(1-k)}{\omega-\frac{1}{2}k^2-{\rm i}\delta}\right]=\Theta(1-k)
\end{equation}
being for small $r_s$ additively corrected by the RPA expression $n_{\rm r}(k)$, which follows from the second term
of Eq. (\ref{c2}) according to
\begin{equation}\label{c4}
n_{\rm r}(k)=\int\limits\frac{d\omega}{2\pi{\rm i}}G_0(k,\omega)\Sigma_{\rm r}(k,\omega)G_0(k,\omega)\ .
\end{equation}
Using $\Sigma_{\rm r}(k,\omega)$ of Eq. (\ref{c1}) it is
\begin{equation}\label{c5} 
n_{\rm r}(k)=\frac{1}{8\pi}\int d^3q \int \frac{d\eta}{2\pi{\rm i}}\ \frac{v^2(\vek q)Q(q,\eta)}{1+v(\vek q)Q(q,\eta)}
\ f(\vek k,\vek q,\eta).
\end{equation}
The case ``2d'' appears, when the ``Yukawa'' term $v(\vek q)Q(q,\eta)$ in the denominator is deleted (`descreening').
The contour integrations are comprised [using $\Theta(k-1)\Theta(1-k)=0$] in
\begin{eqnarray}\label{c6} 
f(\vek k,\vek q,\eta)=\int&\D\frac{d\omega}{2\pi{\rm i}}&
\left[
\frac{\Theta(k-1)}{(\omega-\frac{1}{2}k^2+{\rm i}\delta_1)(\omega-\frac{1}{2}k^2+{\rm i}\delta_2)}+
\frac{\Theta(1-k)}{(\omega-\frac{1}{2}k^2-{\rm i}\delta_1)(\omega-\frac{1}{2}k^2-{\rm i}\delta_2)}
\right]\times \nonumber \\
&\times&
\left[\frac{\Theta(|\vek k+\vek q|-1)}{\omega+\eta-\frac{1}{2}k^2-\vek q(\vek k+\frac{1}{2}\vek q)+{\rm i}\delta }+
\frac{\Theta(1-|\vek k+\vek q|)}{\omega+\eta-\frac{1}{2}k^2-\vek q(\vek k+\frac{1}{2}\vek q)-{\rm i}\delta}\right] .
\nonumber \\
\end{eqnarray}
Only the combinations $\Theta(k-1)\Theta(1-|\vek k+\vek q|)$ with a pole in the upper
$\omega$-plane and $\Theta(1-k)\Theta(|\vek k+\vek q|-1)$ with a pole in the lower plane contribute. 
Next the contour integration along the real $\omega$-axis is closed by half-circles in the upper, respectively
lower plane. Note that $\int\limits_{C_\pm}d\omega/(\omega-\omega_\pm)=\pm2\pi{\rm i}$, where Im $\omega_\pm
\gtrless 0$. The result is
\begin{equation}\label{c7}
f(\vek k,\vek q,\eta)=\frac{\Theta(\vek k, \vek q)}{[\eta- \vek q (\vek k+\frac{1}{2} \vek q)]^2}  
\end{equation}
with $\Theta(\vek k, \vek q)=\pm 1$ for $|\vek k|\gtrless 1, \ |\vek k+\vek q|\lessgtr 1$ and 0 otherwise. This provides with  
the particle-number conservation, because of
\begin{equation}\label{c8}
\left(\int\limits_{|\vek k|>1,|\vek k+\vek q|<1} -\int\limits_{|\vek k|<1,|\vek k+\vek q|>1} \right)
\frac{d^3k}{[\eta-\vek q(\vek k+\frac{1}{2}\vek q)]^2}=0 \quad {\rm or}\quad  \int d^3k\ n_{\rm r}(k)=0.
\end{equation}
Here, with the replacement $\vek k\to -\vek k-\vek q$, the denominator of the second integral transforms to
$[\eta+\vek q(\vek k+\frac{1}{2}\vek q)]^2$. The second integral is identical to the first one. This is because of 
the property $Q(q,-\eta)=Q(q,\eta)$ in the term in front of $f(\vek k,\vek q,\eta)$ in Eq. (\ref{c5}). \\ 

\noindent
Next it is shown, how $t_{\rm r}$ results from $n_{\rm r}(k)$, starting with Eq. (\ref{c5}). 
With the identity (A.5) it can be written as 
\begin{eqnarray}\label{c9}
n_{\rm r}(k)&=&-\frac{1}{4\pi}\int d^3q\int\frac{d\eta}{2\pi {\rm i}}\sum\limits_{n=1}^\infty (-1)^{n+1}v^{n+1}(\vek q)\ Q^n(q,\eta)\
\frac{\delta Q(q,\eta)}{\delta t(\vek k)} \nonumber \\ 
&=&-\frac{1}{4\pi}\int d^3q\int \frac{d\eta}{2\pi{\rm i}}\sum\limits_{n=1}^\infty\frac{(-1)^{n+1}}{n+1}v^{n+1}(\vek q)\
\frac{\delta Q^{n+1}}{\delta t(\vek k)}\ .
\end{eqnarray}
Next $t_{\rm r}=3/(4\pi)\ \int d^3k\ n_{\rm r}(k)\ t(\vek k)$ is combined with the identity (\ref{A6}). It results in
\begin{equation}\label{c10}
t_{\rm r}=\frac{3}{16}\int d^3q \int \frac{d\eta}{2\pi{\rm i}}\sum\limits_{n=1}^\infty (-1)^{n+1}\frac{n}{n+1}
v^{n+1}(\vek q)Q^{n+1}(q,\eta)\ . 
\end{equation}
This is the contribution of the ring-diagram summation to the kinetic energy.
For the total energy contribution $e_{\rm r}$ the potential energy $v_{\rm r}$ is needed. 

\subsection{How $v_{\rm r}$ results from $\Sigma_{\rm r}(k,\omega)$ and $S_{\rm r}(q)$ from $e_{\rm r}$}

\noindent
Equations (\ref{a9}) and (\ref{c1}) inserted into the Galitskii-Migdal formula (\ref{a11}) and the $\omega-$integration performed yields
\begin{eqnarray}\label{c11}
v_{\rm r}&=&\frac{1}{16\pi}\int d^3q \int \frac{d\eta}{2\pi{\rm i}}
\frac{v^2(\vek q)Q(q,\eta)}{1+v(\vek q)Q(q,\eta)}\times \nonumber \\
&\times&\int\frac{3\ d^3k}{4\pi}
\left[\frac{\Theta(|\vek k+\vek q|-1)\Theta(1-k)}{\eta-\vek q(\vek k+\frac{1}{2}\vek q)+{\rm i}\delta}+
\frac{\Theta(1-|\vek k+\vek q|)\Theta(k-1)}{-\eta+\vek q(\vek k+\frac{1}{2}\vek q)+{\rm i}\delta}\right]\ .
\end{eqnarray}
With the definition (\ref{a15}) of $Q(q,\eta)$ it results
\begin{equation}\label{c12} 
v_{\rm r}=-\frac{3}{16\pi}\int d^3q\int \frac{d\eta}{2\pi{\rm i}} \frac{v^2(\vek q)Q^2(q,\eta)}{1+v(\vek q)Q(q,\eta)}\ .   
\end{equation}
The power-series expansion (its 1st term is $v_{\rm 2d}$)
\begin{equation}
v_{\rm r}=-\frac{3}{16\pi}\int d^3q\int\frac{d\eta}{2\pi{\rm i}}\sum\limits_{n=1}^\infty (-1)^{n+1}v^{n+1}(\vek q)Q^{n+1}(q,\eta)\ . 
\nonumber
\end{equation}
makes it better comparable with Eq. (\ref{c10}) for $t_{\rm r}$. Their sum yields
[with $-\frac{n}{n+1}+1=\frac{1}{n+1}$] the well-known RPA expression for the total energy (after Macke)
\begin{equation}\label{c13}
e_{\rm r}=\frac{3}{16\pi}\int d^3q\int\frac{d\eta}{2\pi{\rm i}}[\ln(1+v(\vek q)Q(q,\eta))-v(\vek q)Q(q,\eta)]\ .
\end{equation}
(The 1st term of the power-series expansion gives
 $e_{\rm 2d}$.) Note that $r_sde_{\rm r}/dr_s$ agrees with 
$v_{\rm r}$ of Eq. (\ref{c12}): virial theorem (\ref{a2}). By means of functional derivatives [see Appendix A and Eqs. (\ref{a12})] 
follow the SSF $S_{\rm r}(q)$ and the momentum distribution $n_{\rm r}(k)$. \\ 

\noindent
Indeed, $S_{\rm r}(q)$ results from $e_{\rm r}[t(\vek k),v(\vek q)]$ as 
\begin{equation}\label{c14}
S_{\rm r}(q)= 16\pi\frac{\delta e_{\rm r}}{\delta v(\vek q)}=-3\int \frac{d\eta}{2\pi{\rm i}}\ 
\frac{v(\vek q)Q^2(q,\eta)}{1+v(\vek q)Q(q,\eta)}\ ,
\end{equation}
cf. Fig. 1c, right. If this is multiplied by $v(\vek q)/(16\pi)$ and integrated $\int d^3q$ [according to (1.7)], 
then $v_{\rm r}$ turns out, as it should.  \\ 

\noindent
The analog procedure for $n_{\rm r}(k)$ is
\begin{equation}\label{c15}
n_{\rm r}(k)=\frac{4\pi}{3}\frac{\delta e_{\rm r}}{\delta t(\vek k)}=-\frac{1}{4}\int d^3q \frac{d\eta}{2\pi{\rm i}}\ 
\frac{v^2(\vek q)Q(q,\eta)}{1+v(\vek q)Q(q,\eta)}\frac{\delta Q(q,\eta)}{\delta t(\vek k)}, \quad 
\frac{\delta Q(q,\eta)}{\delta t(\vek k)}=-\frac{1}{2\pi}f(\vek k,\vek q,\eta)\ ,
\end{equation}
in agreement with Eq. (\ref{c5}), as it should. \\

\noindent
In the following it is shown, how Eqs. (\ref{c12}-\ref{c15}) really yield the high-density results
of Macke \cite{Ma}, Gellmann/Brueckner \cite{GB}, Daniel/Vosko \cite{Da}, Kulik \cite{Ku}, and Kimball \cite{Kim3}. 

\subsection{How the complex $Q(q,\eta)$ is replaced by the real $R(q,u)$ and \\ 
how $v_{\rm r}$, $S_{\rm r}(q)$, and $n_{\rm r}(k)$ behave for $r_s\to 0$} 

\noindent
The replacement $\eta\to{\rm i}qu$ turns the contour integration in the RPA expressions (\ref{c12}-\ref{c15}) into one along the real 
axis. Besides it is a useful trick to introduce the velocity $u$ instead of the frequency $\eta$.   
With the replacement $v(\vek q)\to q_c^2/q^2$ they take the form (note $q\ d^3q=2\pi q^2d(q^2)$):
\begin{eqnarray}\label{c16}
{\bf (1)}\hspace{1cm}\qquad e_{\rm r}&=&\frac{3}{8\pi}\int\limits_0^\infty du \int\limits_0^\infty d(q^2)\ q^2
\left[\ln\left(1+\frac{q_c^2}{q^2}R(q,u)\right) -\frac{q_c^2}{q^2}R(q,u)\right]\ , \\
{\bf (2)}\hspace{1cm}\qquad v_{\rm r}&=&-\frac{3q_c^4}{8\pi}\int\limits_0^\infty du \int\limits_0^\infty d(q^2)\ \frac{R^2(q,u)}{q^2+q_c^2R(q,u)}\ , 
\end{eqnarray}
\newpage
\begin{eqnarray}\label{c18}
{\bf (3)}\hspace{4mm}\qquad S_{\rm r}(q)&=&-\frac{3q_c^2}{\pi}\ q\int\limits_0^\infty du\ \frac{R^2(q,u)}{q^2+q_c^2R(q,u)}\ , \\
{\bf (4)}\hspace{4mm}\qquad n_{\rm r}(k)&=&\frac{q_c^4}{8\pi}\int\limits_0^\infty du \int\limits_0^\infty d(q^2)
\frac{R(q,u)}{q^2+q_c^2R(q,u)}\int\limits_{-1}^{+1} d\zeta \frac{\Theta(k,q,\zeta)}{q^2[{\rm i}u-(k\zeta+\frac{1}{2}q)]^2} 
\end{eqnarray}
with $\Theta(k,q,\vek e_k\vek e_q)=\Theta(\vek k,\vek q)$, see (\ref{c7}).
In the following, these four RPA quantities are discussed in detail in the high-density limit $r_s\to 0$. \\

\noindent
{\bf (1)} Let us first consider the {\bf total energy} $e_{\rm r}$ of Eq. (3.16).
The power expansion leads in lowest order to
\begin{equation}\label{c20}
e_{\rm 2d}=-\frac{3q_c^4}{16\pi}\int\limits_0^\infty du \int\limits_{q_0^2}^\infty \frac{d(q^2)}{q^2}\ R^2(q,u) 
=-\frac{3q_c^4}{(8\pi)^2}\int\limits_{q_0}^\infty\frac{dq}{q^2}I(q)\ .
\end{equation}
For the momentum transfer function $I(q)$, to be referred to as Macke function, see (\ref{C11}). Its property $I(q\to 0)\sim q$ 
makes $e_{\rm 2d}$ to diverge for 
$q_0{_> \atop ^{\to}}0$. Vice versa
this flaw of the 2nd-order perturbation theory is rectified by the RPA summation (3.16). How this $e_{\rm r}$ behaves for small 
$r_s$ with the result (\ref{a1}) has been shown by Macke \cite{Ma} and Gell-Mann/Brueckner \cite{GB}. \\

\noindent
{\bf (2)} In the following their method is applied to the {\bf potential energy} $v_{\rm r}$ of Eq. (3.17).
Deleting the term $q_c^2R(q,u)$ in the denominator cancels the RPA partial summation and the diverging expression
\begin{eqnarray}\label{c21}
v_{2{\rm d}}=-\frac{3}{8\pi}q_c^4\int\limits_0^\infty du\int\limits_{q_0^2}^\infty d(q^2)
\frac{R^2(q,u)}{q^2} =-\frac{2\cdot 3q_c^4}{(8\pi)^2}\int\limits_{q_0}^\infty\frac{dq}{q^2}I(q)
\end{eqnarray}
results. It remains to show, how to extract from Eq. (3.17) the constants $c_1$ and $c_2$ of $v_{\rm r}=(\alpha r_s)^2[c_1\ln r_s+c_2+O(r_s)]$ to be
compared with $v_{\rm c}=v_{\rm r}+O(r_s^3)$ of Eq. (\ref{a4}).
Whereas $c_1$ follows from $v_{2{\rm d}}$, $c_2$ results from the peculiar behavior of $v_{\rm r}$ for small
transition momenta $q$. Therefore one can approximate $R(q,u)\approx R_0(u)+\cdots$ and restrict $q$ to $q<q_1$,
where $q_1$ is a small (non-vanishing) momentum: 
\begin{eqnarray}\label{c22}
v_{\rm r}^0&=&-\frac{3}{8\pi}q_c^4\int\limits_0^\infty du\int\limits_0^{q_1^2} d(q^2)
\frac{R_0^2(u)}{q^2+q_{\rm c}^2R_0(u)} \nonumber \\
&=&-\frac{3}{8\pi}q_c^4\int\limits_0^\infty du\ R_0^2(u)
\{\ln[q_1^2+q_{\rm c}^2R_0(u)]-\ln [q_{\rm c}^2R_0(u)]\} \nonumber \\
&=&-\frac{3}{8\pi}q_c^4\int\limits_0^\infty du\ R_0^2(u)
\{[\ln q_1^2+O(r_s)]-\ln [q_{\rm c}^2R_0(u)]\}\ .
\end{eqnarray}
With the constants $a$, $b_{\rm r}'$ defined in Appendix B it is 
\begin{equation}\label{c23}
v_{\rm r}^0=(\alpha r_s)^2 [2a\ln r_s+2(a\ln\frac{4\alpha}{\pi}+b_{\rm r}'-a\ln q_1^2)]+O(r_s^3)\ ,
\end{equation}
thus $c_1=2a$. To find also $c_2$ the difference $\Delta v_{2{\rm d}}=v_{2{\rm d}}-v_{2{\rm d}}^0$ between
the correct 2nd-order term of Eq. (\ref{c22}) and the first term in the expansion of $v_{\rm r}^0$, namely  
\begin{eqnarray}\label{c24}
v_{2{\rm d}}^0=-(\alpha r_s)^2\frac{3}{2\pi^4}\int\limits_{q_0}^{q_1}\frac{dq}{q^2}2\cdot 4\pi q\frac{\pi^3}{3}a
\end{eqnarray}
has to be considered (exploiting the trick of Gell-Mann/Brueckner {\it mutatis mutandi}):
\begin{eqnarray}\label{c25}
\Delta v_{2{\rm d}}&=&-(\alpha r_s)^2\frac{3}{2\pi^4}\left\{\int\limits_{q_0}^\infty\frac{dq}{q^2}I(q)-
\int\limits_{q_0}^{q_1}\frac{dq}{q^2}2\cdot 4\pi q\frac{\pi^3}{3}a\right\}+O(r_s^3) \nonumber \\
&=&(\alpha r_s)^2\left\{-\frac{3}{2\pi^4}\int\limits_{q_0}^\infty \frac{dq}{q^2}\left[I(q)-
\frac{8\pi^4}{3}\frac{a}{q}\Theta(1-q)\right]+4a\int\limits_1^{q_1}\frac{dq}{q}\right\} +O(r_s^3)\ . 
\end{eqnarray}
The first integral does no longer diverge for $q_0\to 0$, therefore one can set $q_0=0$. Besides  
\begin{equation}\label{c26}
\Delta v_{2{\rm d}}=(\alpha r_s)^2 2(b_{2{\rm d}}+a\ln q_1^2)+O(r_s^3)
\end{equation}
shows [for $b_{2{\rm d}}$ see (B.8)], that for $r_s\to 0$ the sum $v_{\rm r}^0+\Delta v_{2{\rm d}}$ does not depend on the arbitrary 
cut-off $q_1$:
\begin{equation}\label{c27}
v_{\rm r}=(\alpha r_s)^2 2 \left[a \ln r_s+(a\ln\frac{4\alpha}{\pi}+b_{\rm r}'+b_{2{\rm d}})\right]+O(r_s^3)\ .
\end{equation}
This has to be compared with Eq. (\ref{a4}).  
Indeed the constant direct term beyond $\ln r_s$ yields 
\begin{equation}\label{c28}
a+2b=2(a\ln\frac{4\alpha}{\pi}+b_{\rm r}'+b_{2{\rm d}})\ ,
\end{equation}
which defines $b$. Its value $b\approx -0.0711$ agrees with what is given in \cite{GB}, as it should. -  The appearence of a 
2nd-order term $\sim r_s^2$ means that - to be consistent - all other terms of the same order contribute to the
small-$r_s$ behavior of $e$. There is only one such term, namely $v_{\rm 2x}$, as treated in Sec. IV. \\ 

\noindent
{\bf (3)} Next the {\bf static structure factor} of Eq. (3.18) is considered. In the lowest order (r$\to$1d) it is  - again with the 
Macke function $I(q)$ of Eq. (\ref{C11}) and with $\omega_{\rm pl}=q_c/\sqrt 3$   
\begin{eqnarray}\label{c29}
S_{\rm 1d}(q)&=&-2\frac{\omega_{\rm pl}^2}{(4\pi/3)^2}\frac{I(q)}{q^2} \quad \curvearrowright \\ 
S_{\rm 1d}(q\to 0)= - 3(1-&\ln 2&)\frac{\omega_{\rm pl}^2}{q}+O(1/q^3)\ , \quad
S_{\rm 1d}(q\to \infty)=-2\frac{\omega_{\rm pl}^2}{q^4}+O(1/q^6)\ . \nonumber 
\end{eqnarray}
At $q=2$, the non-interacting value $S_0(2)=1$ and its
discontinuity jump $\Delta S_0''(2)=3/4$, which arises from the Fermi edge, are reduced by [with $I(2),\Delta I''(2)$ from \cite{Zie5}, 
Eq. (C.2)]
\begin{equation}\label{c30}
S_{\rm 1d}(2)=-(13-16\ln 2)\frac{3}{20}\omega_{\rm pl}^2\ , \quad
\Delta S_{\rm 1d}''(2)=-\left(\frac{3}{4}\right)^2\omega_{\rm pl}^2\ .  
\end{equation}
Note, that $S_{\rm 1x}(q)$ compensates part (ca. half) of $S_{\rm 1d}(q)$, cf. Eq. (\ref{d7}), and note that $S(q)=S_0(q)+
S_{\rm 1d}(q)+S_{\rm 1x}(q)+\cdots$ decreases with increasing $r_s$ for a given value of $q$.
Whereas for $q\gg q_c$ the perturbative treatment $S_{\rm r}(q)=S_{\rm 1d}(q)+O(r_s^2)$ holds (screening effects are not so important for
large momentum transfers $q$),
for $q\ll q_c$ there is a big difference between the `bare' $S_{\rm 1d}(q)$, which behaves unphysically because of $I(q\to 0)\sim q$,
cf. Eq. (\ref{C11}), and its renormalized counterpart $S_{\rm r}(q)$, where the ring-diagram summation ameliorates the above mentioned
flaw of $S_{\rm 1d}(q)$, cf. \cite{Zie3}, Fig. 3.
For small $q$, the approximation $R(q,u)=R_0(u)+\cdots$ is sufficient. So $S_{\rm r}(q)\approx -q_cL(q/q_c)$ with
\begin{equation}\label{c31}
L(y)=\frac{3}{\pi}\ y\int\limits_0^\infty du\ \frac{R_0^2(u)}{y^2+R_0(u)}\quad 
\curvearrowright\quad  L(y\to 0)=\frac{3}{4}y-\frac{\sqrt 3}{2}y^2+\frac{9\sqrt 3}{20}y^4+\cdots ,
\end{equation} 
to be referred to as Kimball function, for its properties cf. \cite{Zie3}. As a consequence, the expression 
\begin{equation}\label{c32}
S_{\rm r}(q\ll q_c)=-\frac{3}{4}q+\frac{q^2}{2\omega_{\rm pl}}-\frac{3}{20}\frac{q^4}{\omega_{\rm pl}^3}\cdots , \quad 
\end{equation} 
(i) eliminates the divergence of $S_{\rm 1d}(q\to 0)$ and (ii) simultaneously
replaces in the sum $S_0(q)+S_{\rm r}(q)$ the linear term of $S_0(q)$ with a quadratic one, which is in agreement
with the plasmon sum rule \cite{Pin,Thom}. Higher-order terms arising from the difference $R(q,u)-R_0(u)=q^2R_1(u)+\cdots$ and from
local field corrections beyond RPA have to kill also the cubic term $-q^3/16$ of $S_0(q)$, to substitute the coefficient of the term 
$\sim q^4/\omega_{\rm pl}^3$ of $S_{\rm r}(q)$ correspondingly, and to add a term $\sim q^5$, such that \cite{GGSB,Iwa}
\begin{equation}\label{c33}
S(q\ll q_c)=\frac{q^2}{2\omega_{\rm pl}}+s_4\frac{q^4}{\omega_{\rm pl}^3}+s_5\frac{q^5}{\omega_{\rm pl}^4}+\cdots, \quad 
s_4<0,\quad \frac{1}{12}<|s_4|<\frac{3}{20}\ .
\end{equation}
A direct consequence of these replacements is the appearance of an inflexion point, which
moves for $r_s\to 0$ towards the origin with $q_{\rm infl}, S_{\rm infl}\sim\omega_{\rm pl}={\sqrt {4\alpha r_s/(3\pi)}}$\ . 
$S_{\rm r}(q)$
of (\ref{c18}) realizes the smooth transition from the $k_{\rm F}(\sim 1/r_s)$-scaling ``far off'' the origin [reasonably approximated 
by $I(q)$ of (\ref{c29})] to the $k_{\rm F}q_c(\sim1/\sqrt r_s)$-scaling near the origin [reasonably approximated by the Kimball-function 
$L(q/q_c)$ of (\ref{c31})]. This transition causes the non-analyticity $v_{\rm r}\sim r_s^2\ln r_s$ \cite{Zie3}. \\   

\noindent
{\bf (4)} Finally the {\bf momentum distribution} (3.19) is considered. 
The expression $q^2f(\vek k,\vek q,{\rm i}qu)$, see (\ref{c7}), is developed in the following way
\begin{equation}\label{c34}
\frac{1}{[{\rm i}u-(k\zeta+\frac{1}{2}q)]^2}=-\frac{1}{[u+{\rm i}(k\zeta+\frac{1}{2})q]^2}=
\frac{\partial}{\partial u}\ \frac{1}{u+{\rm i}(k\zeta+\frac{1}{2}q)}\to\frac{\partial}{\partial u}\
\frac{u}{u^2+(k\zeta+\frac{1}{2}q)^2}\ .
\end{equation}
This allows to write Eq. (3.19) as
\begin{equation}\label{c35}
n_{\rm r}(k)=\frac{\omega_{\rm pl}^4}{(4\pi/3)^2}F_{\rm r}(k) \quad {\rm with} \quad 
F_{\rm r}(k)=\int\limits_0^\infty du\int\frac{d^3q}{q^3}\ \frac{R(q,u)}{q^2+q_c^2R(q,u)}\ F(k,q,u)
\end{equation}
and
\begin{equation}\label{c36}
F(k,q,u)=\frac{\partial}{\partial u}\left(u\int\limits_{-1}^{+1}d\zeta\ \frac{\Theta(k,q,\zeta)}{u^2+(k\zeta+\frac{1}{2}q)^2}\right)\ .
\end{equation}
Next the $\zeta$-integration has to be performed. The boundary conditions $k\gtrless 1$ and
$|\vek k+\vek q|\lessgtr 1$ mean
$k^2+q^2+2kq\zeta\lessgtr 1$ or $\zeta\lessgtr\zeta_0$ with $\zeta_0=(1-k^2-q^2)/(2kq)$.
So we have
\begin{equation}\label{c37}
F(k>1,q,u)=\left.-\frac{1}{k}\frac{k\zeta+\frac{1}{2}q}{(k\zeta+\frac{1}{2}q)^2+u^2}\right|_{-1}^{\zeta_0},\quad
F(k<1,q,u)=\left.\frac{1}{k}\frac{k\zeta+\frac{1}{2}q}{(k\zeta+\frac{1}{2}q)^2+u^2}\right|_{\zeta_0}^{+1}\ .
\end{equation}
For $k>1$ the function $F(k,q,u)$ is non-zero only within a certain stripe of the $k-q-$plane (see Fig. 3), namely
$k-1<q<k+1$ with $\zeta = -1\cdots \zeta_0$ (area I). For $k<1$ the corresponding areas are the triangle
$1-k<q<1+k$ with $\zeta=\zeta_0\cdots +1$ (area II) and the stripe $q>1+k$ with $\zeta=-1\cdots +1$ (area III).
With $k\zeta_0+\frac{1}{2}q=\frac{1-k^2}{2q}$ (in area I and II) it follows
\begin{eqnarray}\label{c38}
F_{\rm I}(k>1,q,u)=+\frac{1}{k}\ \left[\frac{\frac{k^2-1}{2q}}{(\frac{k^2-1}{2q})^2+u^2}-
\frac{k-\frac{1}{2}q}{(k-\frac{1}{2}q)^2+u^2}\right]\quad {\rm for}\quad (k,q) \quad {\rm in} \quad {\rm  I}\ , \nonumber \\
F_{\rm II}(k<1,q,u)=-\frac{1}{k}\ \left[\frac{\frac{1-k^2}{2q}}{(\frac{1-k^2}{2q})^2+u^2}-
\frac{\frac{1}{2}q+k}{(\frac{1}{2}q+k)^2+u^2}\right]\quad {\rm for}\quad (k,q) \quad {\rm in} \quad {\rm II}\ , \nonumber \\
F_{\rm III}(k<1,q,u)=-\frac{1}{k}\ \left[\frac{\frac{1}{2}q-k}{(\frac{1}{2}q-k)^2+u^2}-
\frac{\frac{1}{2}q+k}{(\frac{1}{2}q+k)^2+u^2}\right]\quad {\rm for}\quad  (k,q) \quad {\rm in} \quad {\rm  III}\ .
\end{eqnarray}
This together with (\ref{c35}) is the momentum distribution in the ring-diagram summation \cite{Da,Ku}, see also \cite{Cio0,GGZ}.
For $k=0$ only the last line for III contributes with $F_{\rm III}(0,q,u)=-8(q^2-u^2)/(q^2+u^2)^2$ giving $F_{\rm r}(0)=-4.112335+1.35595\ r_s+\cdots$.
If in the denominator of Eq. (\ref{c35}) the ``Yukawa''-term $q_{\rm c}^2R(q,u)$ is deleted, then 
\begin{equation}\label{c39}
n_{2{\rm d}}(k)=\frac{\omega_{\rm pl}^4}{(4\pi/3)^2}F_{\rm 2d}(k)\ ,\quad 
F_{\rm 2d}(k)=4\pi\int\limits_0^\infty\frac{dq}{q^3}\int\limits_0^\infty du\ R(q,u)\ F(k,q,u)
\end{equation}
arises with $F_{\rm 2d}(k>1)=F_{\rm I}(k)$ and $F_{\rm 2d}(k<1)=F_{\rm II}(k)+F_{\rm III}(k)$. [Note that the definition of 
$F_{\rm 2d}(k)$ with 
${F_{\rm 2d}(k\gtrless 1)}\gtrless 0$ differs from what is used in \cite{Cio0, GGZ}, where $F(k)>0$ for all $k\gtrless 1$].
The function (with $u\sim \tan \varphi$ the $u$-integration is replaced by an angular integration \cite{Cio0}) 
\begin{eqnarray}\label{c40}
F_{\rm 2d}(k>1)=+ \frac{4\pi}{k}\int\limits_{k-1}^{k+1}\frac{dq}{q^3}\int\limits_0^{\pi/2}d\varphi\
[R(q,\frac{k^2-1}{2q}\tan \varphi)-R(q,(k-\frac{1}{2}q)\tan \varphi)]\ , \nonumber \\
F_{\rm 2d}(k<1)= -\frac{4\pi}{k}\int\limits_{1-k}^{1+k}\frac{dq}{q^3}\int\limits_0^{\pi/2}d\varphi\
[R(q,\frac{1-k^2}{2q}\tan \varphi)-R(q,(k+\frac{1}{2}q)\tan \varphi)] \nonumber \\
-\frac{4\pi}{k}\int\limits_{1+k}^{\infty}\frac{dq}{q^3}\int\limits_0^{\pi/2}d\varphi\
[R(q,(\frac{1}{2}q-k)\tan \varphi)-R(q,(\frac{1}{2}q+k)\tan \varphi)]  
\end{eqnarray}
possesses the properties (see also \cite{Cio0,Zie7} and note that $F_{\rm 2x}(k)$ is of the same order, cf. Eq. (\ref{d5}))
\begin{eqnarray}\label{c41}
F_{\rm 2d}(k\to 0)=-\left(4.112335+8.984\ k^2+\cdots\right)\ , \quad
F_{\rm 2d}(k\to \infty) =\frac{1}{2}\frac{(4\pi/3)^2}{k^8}+\cdots\ , \nonumber \\ 
F_{\rm 2d}(k\to 1^{\pm})=\pm\frac{\pi^2}{3}(1-\ln 2)\frac{1}{(k-1)^2}+\cdots . \quad \quad  \quad 
\end{eqnarray}
Consequently, $n_{\rm 2d}(k)$ approximates $n_{\rm r}(k)$ for $k\ll 1$  and $k\gg 1$ (far off the Fermi surface, where the screening 
effect is not so important), but  
it diverges near the Fermi surface square-inversely as $\pm 1/(k-1)^2$ for $k\gtrless 1$.
Vice versa, this divergence is removed through the ring-diagram partial summation with the replacement $q^3\to q[q^2+q_c^2R(q,u)]$
in Eq. (\ref{c39}). For $k$ near the Fermi edge, it holds \cite{Zie3}
\begin{equation}\label{c42}
F_{\rm r}(k\to 1^{\pm})\to \pm \frac{2\pi}{q_{\rm c}^2k^2}G\left(\frac{|k-1|}{q_{\rm c}}\right) \quad {\rm for} \quad 
1\lessgtr k\lessgtr 1\pm\sqrt q_c
\end{equation}
with $G(x)$ being the Kulik function (\ref{B15}). Thus the discontinuity jump for $r_s\to 0$ is described by (the higher-order terms
are different for outside/inside the Fermi surface) 
\begin{equation}\label{c43} 
n_{\rm r}(1^{\pm})=\pm\frac{\omega_{\rm pl}^4}{(4\pi/3)^2}\frac{2\pi}{q_{\rm c}^2}G(0)+\cdots
=\pm \frac{\omega_{\rm pl}^2}{(4\pi/3)^2}\frac{2\pi}{3}G(0)+\cdots= \pm 0.088519\ r_s+\cdots\ . 
\end{equation}
Comparison of (\ref{c41}) with (\ref{c42}) shows that the divergence at $k\to 1^\pm$ is 
eleminated and replaced by the non-analytical behavior of $G(x\to 0)$, see (\ref{B16}). Near the Fermi surface the distribution is
``symmetrical'' and shows a logarithmical ``snuggling'' with infinite slopes:
\begin{eqnarray}
n_{\rm r}(k\to 1^\pm)=n_{\rm r}(1^\pm)\pm\frac{\omega_{\rm pl}}{8}\left(\frac{\sqrt 3\pi}{4}+3\right)|k-1|\ln|k-1|+O(k-1)\ . \nonumber
\end{eqnarray}
$F_{\rm r}(k)$
of (\ref{c35}) realizes the smooth transition from the $k_{\rm F}(\sim 1/r_s)$-scaling ``far off'' the Fermi surface [reasonably 
approximated by $F_{\rm 2d}(k)$ of (\ref{c39})] to the $k_{\rm F}q_c(\sim1/\sqrt r_s)$-scaling near the Fermi surface [reasonably 
approximated by the Kulik-function $G(|k-1|/q_c)$, see (\ref{c42})]. This transition causes the non-analyticity of 
$t_{\rm r}\sim r_s^2\ln r_s$, \cite{Zie3}. - From (\ref{c43}) follows $z_{\rm F}=1-0.177038\ r_s+\cdots$, what is in agreement with the 
Luttinger formula \cite{Lu}, which relates the quasi-particle weight $z_{\rm F}$ directly to the self-energy $\Sigma(k,\omega)$: 
\begin{equation}\label{c44}
z_{\rm F}=\frac{1}{1-\Sigma'(1,\mu)}\ ,\quad \Sigma'(1,\mu)=
\left.{\rm Re}\ \frac{\partial\Sigma(1,\omega)}{\partial \omega}\right|_{\omega=\mu}\ . 
\end{equation}
Indeed, with the ring-diagram approximation (\ref{c1}) it becomes 
\begin{eqnarray}\label{c45}
\Sigma'_{\rm r}(1,\frac{1}{2})&=&
\frac{1}{8\pi}\int d^3q\int \frac{d\eta}{2\pi{\rm i}}\
\frac{v^2(\vek q)Q(q,\eta)}{1+v(\vek q)Q(q,\eta)}
\frac{\partial}{\partial\eta}\D\frac{1}{\eta-{\mbox{\bm $q$}}\cdot(\vek e+\frac{1}{2}\mbox{\bm $q$})\pm{\rm i}\delta}\ , \; 
|\vek e +\vek q|\gtrless 1\ .\nonumber \\ 
\end{eqnarray}
For $r_s\to 0$ it behaves as \cite{Zie0}
\begin{equation}\label{c46}
\Sigma'_{\rm r}(1,\frac{1}{2})=
\frac{\alpha r_s}{\pi^2}\int\limits_0^\infty du\ \frac{R'_0(u)}{\sqrt {R_0(u)}}\ \arctan\frac{1}{u}+\cdots 
\approx -0.177038\ r_s +\cdots\ , 
\end{equation}
in agreement with (\ref{c43}) and (\ref{B17}). (In \cite{Os}, $z_{\rm F}=1-0.12\ r_s+\cdots$ is claimed, instead
of the RPA figure 0.18.) The linear behavior of $z_{\rm F}(r_s)$ is an example for how higher-order 
partial summation may create lower-order terms. Calculations beyond RPA with (\ref{c44}) have been done in \cite{Gel1}. Calculations of 
$z_{\rm F}$ for $r_s\leq 55$ have been done in \cite{Nech}.- The strength of the correlation tail, i.e. 
the relative number of particles [with $k>1$ and using (\ref{B19})] is \cite{Ku}
\begin{equation}\label{c47}
N_{\rm r}=\int\limits_1^\infty d(k^3)\ n_{\rm r}(k)=\omega_{\rm pl}^3\frac{(3/2)^{5/2}}{2\pi^2}\int\limits_0^\infty du\ \frac{R'_0(u)}{\sqrt {R_0(u)}}\ 
u\ln \frac{u^2}{1+u^2} +\cdots \approx 0.05383\ r_s^{3/2}+\cdots 
\end{equation}
\cite{Ku}. The $u$-integral is 1.06252 . How does $n_{\rm 2x}(k)$ change the above results for $z_{\rm F}$ and $N$ ?

\section{Second order: The exchange term $\Sigma_{\rm 2x}$ and its consequences} 
\setcounter{equation}{0}

\noindent
As already mentioned above and stressed by Geldart \cite{Gel1}, for a consistent small-$r_s$ description up to terms 
$\sim r_s^2\ln r_s$ and $\sim r_s^2$ the exchange terms $v_{\rm 2x}$, $e_{\rm 2x}$, $n_{\rm 2x}(k)$, $S_{\rm 1x}(q)$ are needed. 
Here it is shown, how they arise from $\Sigma_{\rm 2x}(k,\omega)$, cf. Fig. 1d, left. \\   

\noindent
The self-energy in the second order of exchange is
\begin{eqnarray}\label{d1}
\Sigma_{2{\rm x}}(k,\omega)&=&\frac{q_c^4}{(8\pi)^2}\int\frac{d^3q_1d^3q_2}{q_1^2q_2^2}\int
\frac{d\eta_1d\eta_2}{(2\pi{\rm i})^2}\times  \\
&\times& G_0(|{\mbox{\bm $k$}}+{\mbox{\bm $q$}}_2|,\omega+\eta_2)G_0(|{\mbox{\bm $k$}}+
{\mbox{\bm $q$}}_1+{\mbox{\bm $q$}}_2|,\omega+\eta_1+\eta_2)G_0(|{\mbox{\bm $k$}}+{\mbox{\bm $q$}}_1|,\omega+\eta_1)\ .
\nonumber
\end{eqnarray}
Use of (\ref{a9}) yields
\begin{eqnarray}\label{d2}
\Sigma_{2{\rm x}}(k,\omega)=-\frac{q_c^4}{(8\pi)^2}\int\frac{d^3q_1d^3q_2}{q_1^2q_2^2}
&\left[\D\frac{\Theta(|{\mbox{\bm $k$}}+{\mbox{\bm $q$}}_1+{\mbox{\bm $q$}}_2|-1)
\Theta(1-|{\mbox{\bm $k$} }+{\mbox{\bm $q$}}_1|)\Theta(1-|{\mbox{\bm $k$}
}+{\mbox{\bm $q$}}_2|)}{\omega-\frac{1}{2}k^2+{\mbox{\bm $q$}}_1\cdot{\mbox{\bm $q$}}_2-
{\rm i} \delta}\right. & \nonumber \\
&+\left.\D\frac{\Theta(1-|{\mbox{\bm $k$}}+{\mbox{\bm $q$}}_1+{\mbox{\bm $q$}}_2|)
\Theta(|{\mbox{\bm $k$} }+{\mbox{\bm $q$}}_1|-1)\Theta(|{\mbox{\bm $k$}
}+{\mbox{\bm $q$}}_2|-1)}{\omega-\frac{1}{2}k^2+{\mbox{\bm $q$}}_1\cdot{\mbox{\bm $q$}}_2+
{\rm i} \delta}\right]& , \nonumber \\
\end{eqnarray}
see also \cite{Zie4}, Eq. (A.5).
This together with (\ref{a9}) used in the Galitskii-Migdal formula (\ref{a11}) gives (cf. Fig. 1d, middle) after the 
$\omega-$integration has been performed
\begin{eqnarray}\label{d3}
v_{2{\rm x}}= -\frac{3q_c^4}{(8\pi)^3}{\rm {Re}}\left[\int\limits _A\frac{d^3kd^3q_1d^3q_2}{q_1^2q_2^2}
\frac{1}{{\vek q}_1\cdot{\vek q}_2+{\rm i}\delta}
+\int\limits_B\frac{d^3kd^3q_1d^3q_2}{q_1^2q_2^2}\frac{1}{{\vek q}_1\cdot(-{\vek q}_2)+{\rm i}\delta}\right]\ .  
\end{eqnarray}
It is easy to show with the help of the substitutions ${\vek q}_1\to{\vek q}_1'$, ${\vek q}_2\to -{\vek q}_2'$,
${\vek k}\to -({\vek k}'+{\vek q}_1')$ that the second term equals the first one. The virial theorem $v_{2{\rm x}}=2e_{2{\rm x}}$ 
gives the 2nd-order exchange energy $e_{\rm 2x}$ to be compared with the 2nd-order direct energy $e_{\rm 2d}$ [see Eq. (\ref{c20})]:
\begin{eqnarray}\label{d4}
e_{2{\rm x}}=- \frac{3q_c^4}{(8\pi)^3}\int\limits_{A}\frac{d^3kd^3q_1d^3q_2}{q_1^2q_2^2}\frac{P}{{\vek q}_1\cdot {\vek q}_2}\ , \quad
e_{2{\rm d}}=+2 \frac{3q_c^4}{(8\pi)^3}\int\limits_{A}\frac{d^3kd^3q_1d^3q_2}{q_1^2q_1^2}\frac{P}{{\vek q}_1\cdot {\vek q}_2}\ . 
\end{eqnarray}
${P}$ means the Cauchy principle value. Note the replacement $1/q_1^2\to 1/q_2^2$ and the addition of a factor $-1/2$, 
when going from the direct term $e_{\rm 2d}$ to the corresponding exchange term $e_{\rm 2x}$.
Having $e_{\rm 2x}$ available, the functions $n_{\rm 2x}(k)$ and $S_{\rm 1x}(q)$ follow by means of the functional derivatives
(\ref{a12}). \\ 

\noindent
Combining Eq. (\ref{a12}) with (\ref{d4}) yields $n_{\rm 2x}(k)=\frac{\omega_{\rm pl}^4}{(4\pi/3)^2}F_{\rm 2x}(k)$ with 
\begin{equation}\label{d5}
 F_{\rm 2x}(k\gtrless 1)=\mp\frac{1}{4}\int\frac{d^3q_1d^3q_2}{q_1^2q_2^2}\frac{1}{({\vek q}_1\cdot {\vek q}_2)^2}, \quad {\rm for} \quad 
 |{\vek k}+{\vek q}_1+{\vek q}_2|\gtrless 1, \quad |{\vek k}+{\vek q}_{1,2}|\lessgtr 1\ .
\end{equation}
For comparison the same procedure with  $e_{\rm 2d}$ yields $n_{\rm 2d}(k)=\frac{\omega_{\rm pl}^4}{(4\pi/3)^2}F_{\rm 2d}(k)$ with 
\begin{equation}\label{d6}
F_{\rm 2d}(k\gtrless 1)=\pm\frac{1}{2}\int\frac{d^3q_1d^3q_2}{q_1^2q_1^2}\frac{1}{({\vek q}_1\cdot {\vek q}_2)^2}\quad
{\rm for} \quad |{\vek k}+{\vek q}_1+{\vek q}_2|\gtrless 1, \quad |{\vek k}+{\vek q}_{1,2}|\lessgtr 1\ .
\end{equation}
It follows from (\ref{C17}), (\ref{C18}), that (\ref{d6}) is equivalent with what arises from (\ref{c5}) for r$\to$2d (`descreening').
Whereas the direct term drives the electrons outside the Fermi surface and decrease the quasi-particle weight $z_{\rm F}$, the exchange 
process draws them back 
inside the Fermi surface and increases $z_{\rm F}$ \cite{Gel1}. Again the boundary conditions enforce $q_{1,2}\to \infty$ for $k\to 
\infty$, so the integral becomes simply $(4\pi/3)^2/k^8$. Hence 
$n_{\rm 2d}(k\to\infty)\to +2\omega_{\rm pl}^4/(4k^8)$ and $n_{\rm 2x}(k\to\infty)\to-\omega_{\rm pl}^4/(4k^8)$ $\curvearrowright$ 
$[n_{\rm r}(k)+n_{\rm 2x}(k)]_{k\to\infty}\to+\omega_{\rm pl}^4/(4k^8)$. Comparison with (\ref{a8}) shows 
$g(0)\approx g_0(0)=1/2$. $F_{\rm 2d,2x}(0)$ are given in (\ref{C22}), (\ref{C23}), integral properties of $F_{\rm 2x}(k)$ are in 
(\ref{C24}), (\ref{C25}). What concerns 
$N_{\rm 2x}$ one should expect $N_{\rm 2x}\approx -\frac{1}{2}N_{\rm r}$, because of 
$n_{\rm 2x}(k\to \infty)=-\frac{1}{2}n_{\rm r}(k\to\infty)$ . Thus $N\approx N_{\rm r}/2+\cdots\approx 0.02691\ r_s^{3/2}+\cdots$ .   \\

\noindent
Combining Eqs. (\ref{a12}) and (\ref{d4}), the 1st-order exchange term of $S(q)$ is (Fig. 1d, right)
\begin{equation}\label{d7}
S_{\rm 1x}(q)=+\left(\frac{\omega_{\rm pl}}{4\pi/3}\right)^2\ \frac{I_{\rm x}(q)}{q^2}\ , \quad 
\frac{I_{\rm x}(q)}{q^2}=-\int\limits_{A} \frac{d^3kd^3q_2}{q_2^2} 
\left .\frac{P}{{\vek q}_1\cdot{\vek q}_2}\right|_{{\vek q}_1\to \vek q}\ .
\end{equation} 
For comparison with $S_{\rm 1d}(q)$ the Macke function $I(k)$ in Eq. (\ref{c29}) is rewritten with (\ref{C11}): 
\begin{equation}\label{d8}
S_{\rm 1d}(q)=-2\left(\frac{\omega_{\rm pl}}{4\pi/3}\right)^2\frac{I(q)}{q^2}\ , \quad 
\frac{I(q)}{q^2}=-\int\limits_{A} \frac{d^3kd^3q_2}{q_1^2} \left .\frac{P}{{\vek q}_1\cdot{\vek q}_2}\right|_{{\vek q}_1\to \vek q}\ .
\end{equation}
They have the asymptotics $S_{\rm 1d,r}(q\to \infty)\to -2\omega_{\rm pl}^2/q^4$ and 
$S_{\rm 1x}(q\to \infty)\to +\omega_{\rm pl}^2/q^4$ $\curvearrowright$
 [$S_{\rm r}(q)+S_{\rm 1x}(q)]_{q\to\infty}=-\omega_{\rm pl}^2/q^4$
\cite{Hol}. Thus the x-term again compensates half of the direct term. Comparison with (\ref{a8}) shows 
$g(0)\approx g_0(0)=1/2$,
as it should up to this order. How does $S_{\rm 1x}(q)$ influence the non-analyticity of $S(q)$ at $q=2$ and the behavior of $S(q)$ at
the origin $q=0$? 

\section{Pair density and short-range correlation}
\setcounter{equation}{0}

\noindent
The SSF $S(q)$ and the PD $g(r)$ are mutually related through the Fourier transforms
\begin{equation}\label{e1}
1-g(r)=\frac{1}{2}\int\limits_0^\infty d(q^3)\ \frac{\sin qr}{qr}[1-S(q)]\ ,\quad  
1-S(q)=\alpha^3\int\limits_0^\infty d(r^3)\ \frac{\sin qr}{qr}\ [1-g(r)]\ .
\end{equation}
$S(0)=0$ expresses the perfect screening sum rule: the normalisation of $1-g(r)$ is $9\pi/4$. About it, the plasmon sum rule says 
$S(q\to 0)=q^2/(2\omega_{\rm pl})+\cdots$. - For $r_s=0$ (ideal Fermi gas) it is [see also Eq. (\ref{b3})] 
\begin{eqnarray}\label{e2}
S_0(q\leq 2)=\frac{3}{2}\frac{q}{2}-\frac{1}2{}\left(\frac{q}{2}\right)^3, \ S_0(q\geq 2)=1\ \curvearrowright\  
g_0(r)=1-\frac{9}{2}\left(\frac{\sin r-r\cos r}{r^3}\right)^2 \leq 1\ . 
\end{eqnarray}
This causes the potential energy in lowest order as $v_{\rm x}=-(3/16)\ q_c^2=-(3/4)^2\ \omega_{\rm pl}^2$. The non-analyticity of $S_0(q)$ at $q=2$, namely
the 2nd-order-derivative jump $\Delta S''_0(2)=3/4$, causes - Fourier transformed - the non-interacting Friedel oscillations of
$g_0(r)$. They are tiny: the 1st minimum at $r\approx 5.76$ is $1-0.0037$. \\



\noindent
Short-range correlation means the behavior of the PD $g(r)$ for $r\ll 1/q_c$, to which belong also (i)
the coalescing cusp and curvature theorems 
\cite{Kim1,Kim4} and (ii) the influence of the on-top PD $g(0)$ on the large-wave-number asymptotics of $n(k)$ and $S(q)$, Eq. (\ref{a8}). \\

\noindent
The relations (\ref{e1}) between the SSF $S(q)$ and the PD $g(r)$ make, that the on-top value $g(0)$ follows
from the normalization (1.7) of $1-S(q)$. The expansion $S(q)=S_0(q)+S_{\rm r}(q)+S_{\rm 1x}(q)+\cdots$ creates corresponding 
on-top terms $g_i(0)$ with $i=0,{\rm r},{\rm 1x},\cdots$. This series starts with $g_0(0)=1/2$ according to (\ref{e2}).
The contribution of $S_{\rm r}(q)$ consists of two parts, a term $S_{\rm 1d}(0)$, linear in $r_s$, and a term $S_{\rm 2r}(q)$, 
logarithmitically non-analytic. The first order direct term is  
\begin{equation}\label{e3}
g_{\rm 1d}(0)= \frac{1}{2}\int\limits_0^\infty d(q^3)\ S_{\rm 1d}(q)=
-\left(\frac{\omega_{\rm pl}}{4\pi/3}\right)^2\int\limits_0^\infty \frac{d(q^3)}{q^2}I(q)
= -2(\pi^2+6\ln 2-3)\frac{\alpha r_s}{5\pi}\approx -0.73167\ r_s\ ,
\end{equation} 
where (\ref{C11}) is used. The factor 2 is killed by $g_{\rm 1x}(0)=-\frac{1}{2}g_{\rm 1d}(0)$. 
So, $g_1(0)=g_{\rm 1d}(0)+g_{\rm 1x}(0)=\frac{1}{2}g_{\rm 1d}(0)\approx -0.3658\ r_s$ \cite{Gel2,Kim3}. Exactly this high-density 
behavior of $g(0)$ results also from the ladder theory
as a method to treat short-range correlation \cite{Cio1,Qi}. The second term $\sim r_s^2\ln r_s$ follows from    
\begin{eqnarray}\label{e4}
S_{\rm 2r}(q)&=&S_{\rm r}(q)-S_{\rm 1d}(q)=\frac{3q_c^4}{\pi}\int\limits_0^\infty du\ \frac{1}{q}\cdot\frac{R^3(q,u)}{q^2+q_c^2R(q,u)}
\quad \curvearrowright \\ 
g_{\rm 2r}(0)=\frac{1}{2}\int\limits_0^\infty d(q^3)\ S_{\rm 2r}(q) 
&=&\frac{9q_c^4}{2\pi}\int\limits_0^\infty du\int\limits_0^\infty qdq\ \frac{R^3(q,u)}{q^2+q_c^2R(q,u)}= 
-2\left(3-\frac{\pi^2}{4}\right)\left(\frac{3\alpha r_s}{2\pi}\right)^2\ln r_s+\cdots, \nonumber
\end{eqnarray}
where $\int\limits_0^\infty du\ R_0^3(u)=\frac{\pi}{8}(3-\frac{\pi^2}{4})$ is used. Again part ($\approx$ half?) of $g_{\rm 2r}(0)$ is 
compensated by a corresponding exchange term. Thus with $x\approx 1/2$ it is \cite{Kim3}
\begin{eqnarray}\label{e5}
g(0)&=&\frac{1}{2}-(\pi^2+6\ln 2 -3)\frac{\alpha}{5\pi}\ r_s-x\ 2\left(3-\frac{\pi^2}{4}\right)\left(\frac{3\alpha}{2\pi}\right)^2 r_s^2 
\ln r_s\ +\cdots \nonumber \\
&=&\frac{1}{2}-0.3658\ r_s-x\ 0.032966\ r_s^2\ln r_s+\cdots\ . 
\end{eqnarray}
The decrease of $g(0)$ with increasing $r_s$ describes the increase of the area between 1 and $S(q)$, the amount of which is $1-g(0)=
1/2+0.3658\ r_s+\cdots$. 

\section{Summary}
\setcounter{equation}{0}

\noindent
Following the procedure of Fig. 2, it is shown for the ground state of the high-density electron gas (as an example), how the 
static 
2-body quantity $S(q)$, the  static structure factor (SSF), follows from the dynamic 1-body quantity $\Sigma(k,\omega)$, the Dyson 
self-energy,
using rigorous theorems as the Migdal formula, the Galitskii-Migdal formula, the generalized Hellmann-Feynman theorem, the virial 
theorem. Along this way  
all the static RPA results are thoroughly revisited and summarized on a unified footing:
 the energy $e$ [Eq. (\ref{c16})] and its components $t$ and $v$, the momentum distribution $n(k)$ [Eq. (3.19)], its behavior for 
$k\to 0$, $k\to \infty$, and at $k\to 1^\pm$ with the discontinuity 
$z_{\rm F}$,  the SSF $S(q)$ [Eq. (3.18)], its behavior at $q\to 0$, and the on-top pair density $g(0)$. Several identities were found, 
e.g. the relation between the SSF $S(q)$ and the Macke function $I(q)$. $S(q)$ and $n(k)$, stemming from the 
2-body density matrix, are simultaneously linked mutually through the self-energy $\Sigma(k,\omega)$, see Fig. 2. An exercise would be 
to perform the ``inverse'' procedure $S(q)\to v\to e\to n(k)$. So far not solved problems:    
To have exactly all terms up to $\sim r_s^2$ and $r_s^2\ln r_s$ available, the functions $I_{\rm x}(q)$ and $F_{\rm 2x}(k)$ of
Eqs. (\ref{C19}) and (\ref{d5}), respectively, have to be calculated. $F_{\rm 2x}(k)$ has to be renormalized. Besides the exchange term,
which compensates part of the direct term $g_{\rm 2r}(0)$, Eq. (\ref{e4}), has to be specified and calculated.

\section*{Acknowledgments}
\noindent
The author is grateful to P. Gori-Giorgi, K. Morawetz, U. Saalmann for discussions and hints and acknowledges P. Fulde for supporting 
this work and thanks Th. M\"uller for technical help.

\begin{appendix}
\section*{Appendix A: Functional derivatives}
\setcounter{equation}{0}
\renewcommand{\theequation}{A.\arabic{equation}}
\noindent
In terms of diagrams, the relations (\ref{a12}) have a simple interpretation according to elementary rules of differentation:
$\delta e/\delta v(\vek q)$ means that in a (vacuum) diagram for $e$ one after the other interaction line has to be cut out 
sucsessively: $\delta v(\vek q')/\delta v(\vek q)=\delta(\vek q-\vek q')$. So
to each $e-$diagram corresponds a set of $S(q)-$diagrams. Because $t(\vek k)$ is in the energy dominator of $G_0(k,\omega)$ (\ref{a9}),
$\delta e/\delta t(\vek k)$ means, that one by one particle-hole line has to be cut up in two sucsessively: 
$\delta G_0(k',\omega)/\delta t(\vek k)=- G_0(k,\omega) \delta(\vek k-\vek k') G_0(k',\omega) $. Thus each $G_0$ has to be 
replaced by $-G_0G_0$. The minus sign is because a closed loop is broken and the 2 $G_0$'s are the outer ends of an $n(k)-$diagram, 
corresponding to $G_0\Sigma G_0$, see (\ref{c4}), (\ref{C16}). \\

\noindent
The RPA energy $e_{\rm r}$ of (\ref{c13}) is of the kind
\begin{equation}\label{A1} 
e_{\rm r}[t(\vek k), v(\vek q)]=\int d^3q\int d\eta\ f(Q(q,\eta), v(\vek q))  
\end{equation}
with $Q(q,\eta)$ being a functional of $t(\vek k)$. From the derivative
\begin{equation}\label{A2}
\frac{\delta e_{\rm r}}{\delta v(\vek q)}=\int d^3q'\int d\eta\ \left .\frac{\partial f(Q(q',\eta),v)}{\partial v}
\right|_{v\to v({\vek q}')}
\delta (\vek q'-\vek q)=\int d\eta \left .\frac{\partial f(Q(q,\eta),v)}{\partial v}\right|_{v\to v(\vek q)}
\end{equation}
follows $S_{\rm r}(q)$ of Eq. (\ref{c14}), as it should. From the derivative
\begin{equation}\label{A3} 
\frac{\delta e_{\rm r}}{\delta t(\vek k)}= \int d^3q\int d\eta\left . \frac{\partial f(Q,v(\vek q)}{\partial Q}\right|_{Q\to Q(q,\eta)}
\frac{\delta Q(q,\eta,[t(\vek k)])}{\delta t(\vek k)}\ .
\end{equation}
follows $n_{\rm r}(k)$ as soon as the task $\delta Q/ \delta t$ is solved. This is done with the help of 
\begin{equation}\label{A4}
\frac{\delta}{\delta t(\vek k)}\int\frac{d^3k'}{t({\vek k}'+\vek q)-t({\vek k}')\pm \eta-{\rm i}\delta} =-\int d^3k'
\frac{\delta({\vek k}'+\vek q-\vek k)-\delta({\vek k}'-\vek k)}{[t({\vek k}'+\vek q)-t({\vek k}')\pm \eta-{\rm i}\delta]^2}\ .
\end{equation}
This result combined with the definition (\ref{a15}) of $Q(q,\eta)$ yields
\begin{equation}\label{A5}
\frac{\delta Q(q,\eta)}{\delta t(\vek k)}=-\frac{1}{2\pi}f(\vek k,\vek q,\eta)\ .
\end{equation}
This identity used in (\ref{a12}) gives the expression (\ref{c15}), which has been derived from the RPA-self-energy 
$\Sigma_{\rm r}(k,\omega)$. \\

\noindent
To derive the identity
\begin{equation}\label{A6}
\int d^3k\ t(\vek k)\frac{\delta}{\delta t(\vek k)}\int d\eta\ Q^{n+1}(q,\eta)=-n\int d\eta\ Q^{n+1}(q,\eta)\ .
\end{equation}
one must have in mind, that $Q(q,\eta)$ has the structure $\hat Q f(\eta)$ with 
$f(\eta)=1/(\tau-\eta-{\rm i}\delta)+1/(\tau+\eta-{\rm i}\delta)$ with the excitation energy $\tau=t(\vek k+\vek q)-t(\vek k)$.
After the contour integration (only) $n$ factors $f_i(\tau_{n+1})$ remain:
\begin{equation}\label {A7}
\int d\eta\ Q^{n+1}(q,\eta)= \hat Q_1\cdots \hat Q_{n+1}\int d\eta\ f_1(\eta)\cdots f_{n+1}(\eta)=
\hat Q_1\cdots \hat Q_{n+1} f_1(\tau_{n+1})\cdots f_n(\tau_{n+1})\ .
\end{equation} 
Now the operation $\int d^3k\ t(\vek k)\delta/\delta t(\vek k)$ counts these remaining $n$ factors. Finally the energy {\it denominators}
$1/(\tau_i+\tau_{n+1})$ cause the minus sign.

\section*{Appendix B: The function $R(q,u)$ and characteristic RPA constants}
\setcounter{equation}{0}
\renewcommand{\theequation}{B.\arabic{equation}}
\noindent
From the particle-hole propagator (\ref{a15}) follows the real function $R(q,u)=Q(q,{\rm i}qu)$ with 
\begin{eqnarray}\label{B1}
R(q,u)=\frac{1}{2} \left[ 1+\frac{1+u^2-\frac{q^2}{4}}{2q}\ln\frac{(\frac{q}{2}+1)^2+u^2}{(\frac{q}{2}-1)^2+u^2}
- u\left(\arctan\frac{1+\frac{q}{2}}{u}+\arctan\frac{1-\frac{q}{2}}{u}\right) \right] .
\end{eqnarray}
It has the $q$-expansion $R(q,u)=R_0(u)+q^2R_1(u)+\cdots$ with
\begin{eqnarray}\label{B2}
R_0(u)=1-u\arctan \frac{1}{u}\ ,\quad R_1(u)=-\frac{1}{12(1+u^2)^2}\ , \ \cdots\ .
\end{eqnarray}
Its derivative with respect to $u$ is
\begin{equation}\label{B3}
R'(q,u)= -\frac{1}{2}\left( \arctan\frac{1+\frac{q}{2}}{u}+\arctan\frac{1-\frac{q}{2}}{u} \right)+
\frac{u}{2q}\ln\frac{(\frac{q}{2}+1)^2+u^2}{(\frac{q}{2}-1)^2+u^2}
\end{equation}
with the $q$-expansion $R'(q,u)=R_0'(u)+q^2R_1'(u)+\cdots$ with
\begin{equation}\label{B4}
R_0'(u)=\frac{u}{1+u^2}-\arctan\frac{1}{u}\ , \quad R_1'(u)=\frac{u}{3(1+u^2)^3}, \ \cdots , \quad
R'(q\to \infty,u)=-\frac{32u}{3q^4}+\cdots\ .
\end{equation}
For the integral $\int\limits_0^\infty du\ R^2(q,u)$ see (\ref{C12}). Important RPA constants are:
\begin{eqnarray}
a&=&\frac{3}{\pi^3}\int\limits_0^\infty du\; R_0^2(u)=\frac{1-\ln 2}{\pi^2}\approx 0.031091\label{B5} , \\
b_{\rm r}'&=&\frac{3}{\pi^3}\int\limits_0^\infty du \; R_0^2(u)\ln R_0(u)\approx -0.0171202\label{B6} , \\
b_{\rm r}&=& b_{\rm r}'+a\left(\ln \frac{4\alpha}{\pi}-\frac{1}{2}\right) \approx -0.045423\label{B7} , \\
b_{2{\rm d}}&=&-\frac{3}{4\pi^4}\int\limits_0^1 \frac{dq}{q}\left[\frac{I(q)}{q} -\frac{8\pi^4}{3}a\
\Theta(1-q)\right] \nonumber \\
&=& \frac{1}{4}+\frac{1}{\pi^2}\left[-\frac{11}{6}-\frac{8}{3}\ln2+2(\ln2)^2\right]\approx -0.025677\label{B8} , \\
b_{2{\rm x}}&=&\frac{1}{6}\ln 2-\frac{3}{4}\frac{\zeta(3)}{\pi^2}\approx +0.02418\label{B9} .
\end{eqnarray}
Here is a list of integrals:
\begin{eqnarray}
\int\limits_0^\infty du\; R_0^2(u)&=&\frac{\pi^3}{3}a\approx 0.321336,\quad
\int\limits_0^\infty du \; R_0^2(u)\ln R_0(u)\approx -0.176945,\label{A12}  \\
\int\limits_0^\infty du\ \frac{R_0(u)}{1+u^2}&=&\frac{\pi^3}{2}a\approx 0.482003, \quad
\int\limits_0^\infty du\ \frac{R_0(u)\ln R_0(u)}{1+u^2}\approx-0.345751,\label{A13}
\end{eqnarray}
\begin{equation}\label{B12}
\int\limits_0^\infty du\; \frac{R'_0(u)}{R_0(u)}\arctan \frac{1}{u}\approx -3.353337\ , \quad
\int\limits_0^\infty du\; \frac{R''_0(u)}{R_0(u)}\arctan \frac{1}{u}\approx  4.581817 .
\end{equation}
\begin{equation}\label{B13}
\frac{2}{\pi^3}\int\limits_0^\infty du\ R_0(u)\ln R_0(u)\left[\frac{1}{1+u^2}-\frac{3}{2}R_0(u)\right]=-\frac{1}{6}a\ ,
\end{equation}
\begin{equation}\label{B14}
\frac{4}{\pi}\int\limits_0^\infty du\ \left[\frac{uR_0'(u)}{4(1+u^2)^2}-\left(\arctan\frac{1}{u}\right)R_1'(u)\right]=
-\frac{1}{6} \ .
\end{equation}
The Kulik function $G(x)$ approximates $F_{\rm r}(k)$ for $k\approx 1$, see (\ref{c42}). It follows from $R_0(u)$ according to 
\begin{equation}\label{B15}
G(x)=\left .\int\limits_0^\infty du\ \frac{R'_0(u)}{R_0(u)}\cdot\frac{u}{u+y}\cdot\frac{R_0(u)-R_0(y)}{u-y}\right|_{y=x/\sqrt {R_0(u)}}\ .
\end{equation}
It has the small-$x$ behavior
\begin{equation}\label{B16}
G(x\ll 1)=G(0)+ \left[\pi\left(\frac{\pi}{4}+\sqrt 3\right)x+O(x^2)\right]\ln x+O(x)
\end{equation}
with 
\begin{equation}\label{B17}
G(0)=-\int\limits_0^\infty du\ \frac{R'_0(u)}{R_0(u)}\arctan{\frac{1}{u}}\approx 3.353\ 337\ .
\end{equation}
The coefficient of $x\ln x$ is 7.908\ 799 (the Kulik number). $G(x)$ has the large $x$-behavior
\begin{equation}\label{B18} 
G(x\gg 1)=\frac{\pi}{6}(1-\ln 2)\frac{1}{x^2}\approx 0.160\ 668\frac{1}{x^2} .
\end{equation} 
The Kulik function is shown in \cite{GGZ}, Fig. 1. \\

\noindent
The relative number of particles, $N_{\rm r}$ of (\ref{c47}), follows from $F_{\rm I}(k)$ of (\ref{c38}) as 
\begin{equation}\label{B19}
N_{\rm r}=-\frac{3q_c^4}{4\pi}\int\limits_0^\infty du \int_{\rm I} q dq\ k dk\
\frac{R'(q,u)}{[q^2+q_c^2R(q,u)]^2}\left[\arctan\frac{u}{(k^2-1)/(2q)}-\arctan\frac{u}{k-q/2}\right]\ . 
\end{equation}
In lowest order the decisive region is $k\gtrapprox 1$ and $0<q\lessapprox \delta$:
\begin{equation}\label{B20}
N_{\rm r}=-\frac{3}{4\pi}q_c^4\int\limits_0^\infty du \int\limits_0^\delta q dq\
\frac{R'(q,u)}{[q^2+q_c^2R(q,u)]^2}\int\limits_1^{1+q} k dk\left[\arctan\frac{u}{(k^2-1)/(2q)}-\arctan\frac{u}{k-q/2}\right]\ .
\end{equation}
The $k$-integral yields (expanded in powers of $q$)
\begin{equation}
\frac{1}{2}\left(\arctan\frac{1}{u}- \arctan u-\frac{\pi}{2}\right)+\frac{1}{2}u\ln \left(1+\frac{1}{u^2}\right)\ q+\cdots\ . \nonumber
\end{equation}
The 1st ($q$-independent) term does not contribute, because of $\int\limits_0^\infty du\ R'_{0}(u)(\cdots)=0$. The $q$-integral 
yields in the limit $r_s\to 0$ the expression $\pi/(4q_c\sqrt {R_0(u)})$, giving (\ref{c47}).  

\section*{Appendix C: Identities, Macke function, momentum distribution at $k=0$ }
\setcounter{equation}{0}
\renewcommand{\theequation}{C.\arabic{equation}}
\noindent
For the regions of the wave-number integrations the following short hands are introduced
\begin{eqnarray}
A=({k<1,\ |{\vek k}+{\vek q}_1+{\vek q}_2|<1,\ |\vek k+{\vek q}_{1,2}|>1}), \quad \tilde A =(k_{1,2}<1,\ |{\vek k}_{1,2}+\vek q|>1)\ , \\
B=({k>1,\ |{\vek k}+{\vek q}_1+{\vek q}_2|>1,\ |\vek k+{\vek q}_{1,2}|<1}), \quad \tilde B =(k_{1,2}>1,\ |{\vek k}_{1,2}+\vek q|<1)\ . 
\end{eqnarray}
For the Heisenberg energy $e_{\rm 2d}$, the Onsager energy $e_{\rm 2x}$, for the SSF's $S_{\rm 1d,1x}(q)$ and for the momentum 
distributions $n_{\rm 2d,2x}(k)$ the following equivalent writings exist: 
\begin{equation}\label{C3}
e_{\rm 2d}:\quad \int\limits_A\frac{d^3kd^3q_1d^3q_2}{q_1^2q_1^2}\frac{P}{{\vek q}_1\cdot{\vek q}_2}= 
-\int\limits_{\tilde A}\frac{d^3qd^3k_1d^3k_2}{q^2q^2}\frac{P}{\vek q\cdot({\vek k}_1+{\vek k}_2+{\vek q})}\ ,
\end{equation}
\begin{equation}\label{C4}
e_{\rm 2x}:\quad \int\limits_A\frac{d^3kd^3q_1d^3q_2}{q_1^2q_2^2}\frac{P}{{\vek q}_1\cdot{\vek q}_2}= 
-\int\limits_{\tilde A}\frac{d^3qd^3k_1d^3k_2}{q^2({\vek k}_1+{\vek k}_2+{\vek q})^2}
\frac{P}{\vek q\cdot({\vek k}_1+{\vek k}_2+{\vek q})}\ ,
\end{equation}
\begin{equation}\label{C5}
S_{\rm 1d}(q):\quad \quad \left . \int\limits_A d^3kd^3q_2\frac{P}{{\vek q}_1\cdot{\vek q}_2}\right |_{{\vek q}_1\to \vek q}=
-\int\limits_{\tilde A} d^3k_1d^3k_2
\frac{P}{\vek q\cdot({\vek k}_1+{\vek k}_2+{\vek q})}\ ,
\end{equation}
\begin{equation}\label{C6}
S_{\rm 1x}(q):\quad \left . \int\limits_A \frac{d^3kd^3q_2}{q_2^2}\frac{P}{{\vek q}_1\cdot{\vek q}_2}\right |_{{\vek q}_1\to \vek q}=
-\int\limits_{\tilde A} \frac{d^3k_1d^3k_2}{({\vek k}_1+{\vek k}_2+{\vek q})^2}
\frac{P}{\vek q\cdot({\vek k}_1+{\vek k}_2+{\vek q})}\ ,
\end{equation}
\begin{equation}\label{C7}
n_{\rm 2d}(k<1):\quad \int\limits_A\frac{d^3q_1d^3q_2}{q_1^2q_1^2}\frac{1}{({\vek q}_1\cdot{\vek q}_2)^2}=\left .
\int\limits_{\tilde A}\frac{d^3qd^3k_2}{q^2q^2}\frac{P}{[\vek q\cdot({\vek k}_1+{\vek k}_2+{\vek q})]^2}\right|_{{\vek k}_1\to\vek k}\ , 
\end{equation}
\begin{equation}\label{C8}
n_{\rm 2d}(k>1):\quad \int\limits_B\frac{d^3q_1d^3q_2}{q_1^2q_1^2}\frac{1}{({\vek q}_1\cdot{\vek q}_2)^2}=\left .
\int\limits_{\tilde B}\frac{d^3qd^3k_2}{q^2q^2}\frac{P}{[\vek q\cdot({\vek k}_1+{\vek k}_2+{\vek q})]^2}\right|_{{\vek k}_1\to\vek k}\ ,
\end{equation}
\begin{equation}\label{C9}
n_{\rm 2x}(k<1):\quad \int\limits_A \frac{d^3q_1d^3q_2}{q_1^2q_2^2} \frac{1}{({\vek q}_1\cdot{\vek q}_2)^2}=\left .
\int\limits_{\tilde A} \frac{d^3qd^3k_2}{q^2({\vek k}_1+{\vek k}_2+{\vek q})^2} \frac{P}{[\vek q\cdot({\vek k}_1+{\vek k}_2+{\vek q})]^2}
\right|_{{\vek k}_1\to\vek k}\ ,
\end{equation}
\begin{equation}\label{C10}
n_{\rm 2x}(k>1):\quad \int\limits_B\frac{d^3q_1d^3q_2}{q_1^2q_2^2}\frac{1}{({\vek q}_1\cdot{\vek q}_2)^2}=\left .
\int\limits_{\tilde B}\frac{d^3qd^3k_2}{q^2({\vek k}_1+{\vek k}_2+{\vek q})^2}
\frac{P}{[\vek q\cdot({\vek k}_1+{\vek k}_2+{\vek q})]^2}\right|_{{\vek k}_1\to\vek k}\ . 
\end{equation}
These equivalences are consequences of the invariance, when changing the integration variables according to 
\begin{equation}
{\vek k} \longleftrightarrow {\vek k}_1\ , \quad {\vek q}_1\longleftrightarrow \vek q\ , \quad 
{\vek q}_2\longleftrightarrow -({\vek k}_1+{\vek k}_2+{\vek q})\ . \nonumber
\end{equation} 

\noindent
Note that Eq. (\ref{C5}) defines a function $I(q)$, which has been calculated explicitly by Macke: 
\begin{eqnarray}\label{C11}
I(q)=-\left.\int\limits_A d^3kd^3q_2\ \frac{P}{{\vek q_1}\cdot{\vek q_2}}\right|_{\vek q_1\to \vek q}\quad {\rm or}\quad  
I(q)&=&+\int\limits_{\tilde A} d^3k_1d^3k_2\ \frac{P}{\vek q\cdot({\vek k}_1+{\vek k}_2+{\vek q})}\quad  \curvearrowright \nonumber   \\  
 I(q\to 0)=\frac{8\pi^4}{3}aq+\cdots\ , \; I(q\to \infty)=\frac{(4\pi/3)^2}{q^2}&+&\cdots\ ,\; 
\int\limits_0^\infty dq\ I(q)=\frac{8\pi^2}{45}(\pi^2+6\ln 2 -3)\ .  \nonumber \\
\end{eqnarray}
The linear behavier for $q\to 0$ causes the Heisenberg divergence of $e_{\rm 2d}$, the large-$q$ asymptotics makes 
$S(q\to\infty)\sim 1/q^4$. 
At $q=2$, $I(q)$ has an inflexion point and  $I''(q)$ has the jump discontinuity  $\Delta I''(2)=2\pi^2$, besides 
$I(q)=\cdots +c_{\pm}\frac{\pi^2}{3} (q-2)^3\ln |q-2|+\cdots$ with $c_+=+1$ for $q>2$ and $c_-=-1/2$ for $q<2$\ . The non-analyticity 
at $q=2$ influences the Friedel oscillations of $g(r\to \infty)$. For further properties see \cite{Zie5}, 
Appendix C, \cite{Ho}. The identity
\begin{equation}\label{C12}
8\pi q \int\limits_0^\infty du\ R^2(q,u)=I(q)
\end{equation} 
is proven as it follows. On the one hand it is (with $\eta={\rm i}qu$ and $R(q,u)=Q(q,{\rm i}qu))$
\begin{equation}\label{C13}
8\pi^2\int\frac{d\eta}{2\pi{\rm i}}Q^2(q,\eta)=8\pi q\int\limits_0^\infty du\ R^2(q,u)\ . 
\end{equation} 
On the other hand the lhs of (\ref{C13}) - with the definition (\ref{a15}) of $Q(q,\eta)$ - can be written as [with the particle-hole 
excitation energy $\tau_i=t(\vek k_i+\vek q)-t(\vek k_i)$]
\begin{equation}\label{C14}
\frac{8\pi^2}{(4\pi)^2}\int\limits_{\tilde A}d^3k_1d^3k_2\int\frac{d\eta}{2\pi{\rm i}}
\left[\frac{1}{\tau_1-\eta-{\rm i}\delta}\cdot\frac{1}{\tau_2+\eta-{\rm i}\delta}+
\frac{1}{\tau_1+\eta-{\rm i}\delta}\cdot\frac{1}{\tau_2-\eta-{\rm i}\delta}\right]\ .
\end{equation} 
The frequency integration is performed as a contour integration with the result
\begin{equation}\label{C15}
8\pi q\int\limits_0^\infty du\ R^2(q,u)=
\int\limits_{\tilde A}d^3k_1d^3k_2\frac{P}{[t({\vek k}_1+{\vek q})-t({\vek k}_1)]+[t({\vek k}_2+{\vek q})-t({\vek k}_2)]}\ .
\end{equation}  
With $t(\vek k)=k^2/2$ the Macke function $I(q)$ of (\ref{C11}) turns out, qed. \\ 

\noindent
With the help of the functional derivative
\begin{equation}\label{C16}
\frac{\delta}{\delta t(\vek k)}\frac{1}{{\vek q} \cdot ({\vek k}_1+{\vek k}_2+{\vek q})}
=-\frac{\delta({\vek k}_1+\vek q-\vek k)-\delta({\vek k}_1-\vek k)+\delta({\vek k}_2+\vek q-\vek k)-\delta({\vek k}_2-\vek k)}
{[\vek q\cdot({\vek k}_1+{\vek k}_2+{\vek q})]^2}\ ,
\end{equation}
it is easy to show
\begin{equation}\label{C17}
\frac{\delta I(q)}{\delta t(\vek k_1)}=\mp 2\int\frac{d^3k_2}{[\vek q \cdot(\vek k_1+\vek k_2+\vek q)]^2}\quad {\rm for}\quad 
k_{1,2}\gtrless 1,\; |\vek k_{1,2}+\vek q|\lessgtr 1
\end{equation}
With the help of (\ref{C12}), (\ref{C13}), and (\ref{A5}) it is again easy to show also
\begin{equation}\label{C18}
\frac{\delta I(q)}{\delta t(\vek k_1)}=-8 q R(q,u) f(\vek k_1, \vek q, {\rm i}qu)\ .
\end{equation}
(\ref{C17}) and (\ref{C18}) together with (\ref{C7}), (\ref{C8}) show the equivalence of (\ref{c39}) [$n_{\rm 2d}(k)$ derived from the 
self-energy $\Sigma_{\rm 2d}(k,\omega)$] with (\ref{d6}) [$n_{\rm 2d}(k)$ derived from $e_{\rm 2d}$]. \\ 

\noindent
${\bf S_{\rm 1x}(q):}$ The exchange counterpart of $I(q)$, namely
\begin{equation}\label{C19}
I_{\rm x}(q)=-q^2\int\limits_{A} \frac{d^3kd^3q_2}{q_2^2}
\left .\frac{P}{{\vek q}_1\cdot{\vek q}_2}\right|_{{\vek q}_1\to \vek q} \quad {\rm or}\quad
I_{\rm x}(q)=+\int\limits_{\tilde A} \frac{d^3k_1d^3k_2}{({\vek k}_1+{\vek k}_2+{\vek q})^2}
\frac{P}{\vek q\cdot({\vek k}_1+{\vek k}_2+{\vek q})} 
\end{equation}
has the asymptotics $I_{\rm x}(q\to \infty)\to (4\pi/3)^2/q^2$, the same as $I(q)$. The properties
\begin{equation}\label{C20}
\int\limits_0^\infty dq\ I_{\rm x}(q)=\int\limits_0^\infty dq\ I(q)\quad {\rm and}\quad 
\int\limits_0^\infty \frac{dq}{q^2}I_{\rm x}(q)= \frac{8\pi^4}{3} \left[\frac{1}{6}\ln 2-3\frac{\zeta(3)}{(2\pi)^2}\right]
\end{equation}
follow from 
\begin{equation}\label{C21}
\int\limits_A \frac{d^3kd^3q_1d^3q_2}{q_2^2\ \vek q_1\cdot \vek q_2}=
\int\limits_A \frac{d^3kd^3q_1d^3q_2}{q_1^2\ \vek q_1\cdot \vek q_2}\quad {\rm and}\quad  
 \frac{q_c^2}{4}\int\limits_0^\infty dq\ S_{\rm 1x}(q)=v_{\rm 2x}=2 b_{\rm 2x}(\alpha r_s)^2\ ,
\end{equation}
 respectively. One may guess, that $I_{\rm x}(q\to 0)$ starts with $\sim q^6$. 
This would contribute to $s_4$ of Eq. (\ref{c33}). \\

\noindent
${\bf n_{\rm 2d,2x}(k):}$ Here $F_{\rm 2d,2x}(0)$ are calculated starting with (\ref{d5}), (\ref{d6}) for $k=0$. The condition $|\vek q_1+\vek q_2|<1$ means
with $\zeta=\cos \varangle (\vek q_1\cdot\vek q_2 )$ the restrictions $\zeta=-1\cdots -(1-q_1^2-q_2^2)/(2q_1q_2)$ and $q_1-1<q_2<q_1+1$:
\begin{equation}\label{C22}
F_{\rm 2d}(0)=-4\pi^2\left(\int\limits_1^2\frac{dq_1}{q_1^4}\int\limits_1^{q_1+1}dq_2+
\int\limits_2^\infty\frac{dq_1}{q_1^4}\int\limits_{q_1-1}^{q_1+1}dq_2\right)\frac{1-(q_1-q_2)^2}{q_1^2+q_2^2-1}=-\frac{5\pi^2}{12}\ .
\end{equation}
For the corresponding exchange term it is similarly
\begin{equation}\label{C23}
F_{\rm 2x}(0)=+2\pi^2\left(\int\limits_1^2\frac{dq_1}{q_1^2}\int\limits_1^{q_1+1}\frac{dq_2}{q_2^2}+
\int\limits_2^\infty\frac{dq_1}{q_1^2}\int\limits_{q_1-1}^{q_1+1}\frac{dq_2}{q_2^2}\right)\frac{1-(q_1-q_2)^2}{q_1^2+q_2^2-1}=
+2\pi^2(1-\ln 2)^2\ ,
\end{equation}
thus $F_{\rm 2}(0)\approx-4.1123+1.8586\approx-2.2537$ and $n_{\rm 2}(0)\approx-0.0063\ r_s^2$. - $F_{\rm 2d,2x}(k)$ have the asymptotics
$\to -\frac{(4\pi/3)^2}{4k^8}$. The particle-number conservation
\begin{equation}\label{C24}
\left(\int\limits_{B}-\int\limits_{A}\right)dk\ k^2 \int\frac{d^3q_1d^3q_2}{q_1^2q_2^2}\frac{1}{(\vek q_1\cdot \vek q_2)^2}=0
\end{equation}
easily follows from the replacements in the second integral $\vek k\to -\vek k-q_1$ and $\vek q_2\to -\vek q_2$. 
The property 
\begin{equation}\label{C25}
\left(\int\limits_{B}-\int\limits_{A}\right)dk\ k^4 \int\frac{d^3q_1d^3q_2}{q_1^2q_2^2}\frac{1}{(\vek q_1\cdot \vek q_2)^2}=
\frac{8\pi^4}{9}\left[\frac{1}{6}-3\frac{\zeta(3)}{(2\pi)^2}\right]
\end{equation}
follows from 
\begin{equation}\label{C26}
\frac{\omega_{\rm pl}^4}{(4\pi/3)^2}\int\limits_0^\infty d(k^3)F_{\rm 2x}(k)\frac{k^2}{2}=t_{\rm 2x}=-b_{\rm 2x}(\alpha r_s)^2\ . 
\end{equation}
How behaves $F_{\rm 2x}(k)$ near the Fermi edge ? It diverges, but weaker than $F_{\rm 2d}(k)$, namely logarithmically,
$F_{\rm 2x}(k\to 1^\pm)\sim \pm\ln |k-1|$. It is not strong enough to make the corresponding (Onsager) energy $t_{\rm 2x}$ divergent:
\begin{equation}\label{C27}
\left .\left(\int\limits_1^{1+\Delta}-\int\limits_{1-\Delta}^1\right) dk\ k^4\ \ln |k-1|\right|_{\Delta\to 0} \sim
\Delta^2\ln\Delta\to 0\ ,
\end{equation}
saying that the lhs of Eq. (\ref{C26}) exists.
The divergence of $F_{\rm 2x}(k)$ is eleminated by the renormalization of at least one interaction line.

\end{appendix}

\begin{center}
{\bf Figures}
\end{center}

\begin{figure}[h!] 
\begin{center}
\rotatebox{0}{\resizebox{75mm}{!}{\includegraphics{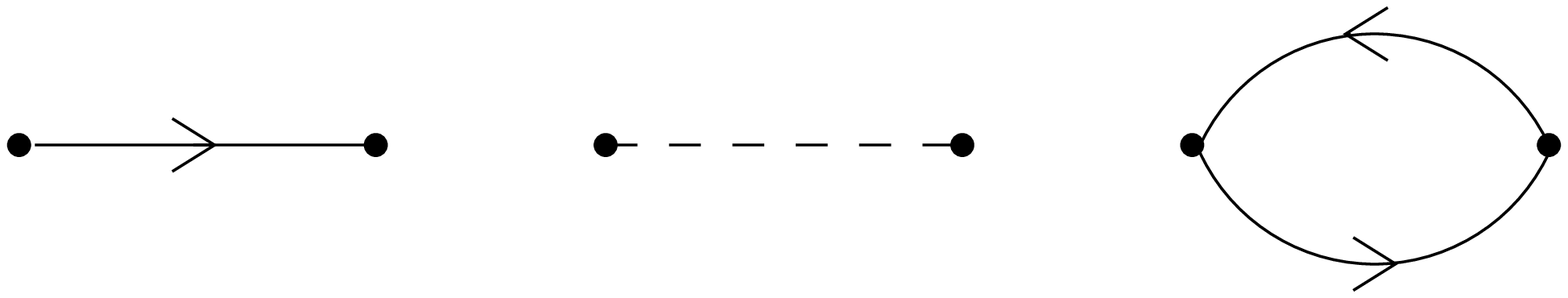}}}
\end{center}
\end{figure} 
\noindent
FIG. 1a: Feynman diagrams for the one-body Green's function of the ideal Fermi gas $G_0(k,\omega)$, 
the bare Coulomb repulsion $v_0(q)=\pi^2q_c^2/q^2$, and the RPA polarization propagator $Q(q,\eta)$ as defined in 
Eqs. (\ref{a13})-(\ref{a15}). 

\begin{figure}[h!]
\begin{center}
\rotatebox{0}{\resizebox{100mm}{!}{\includegraphics{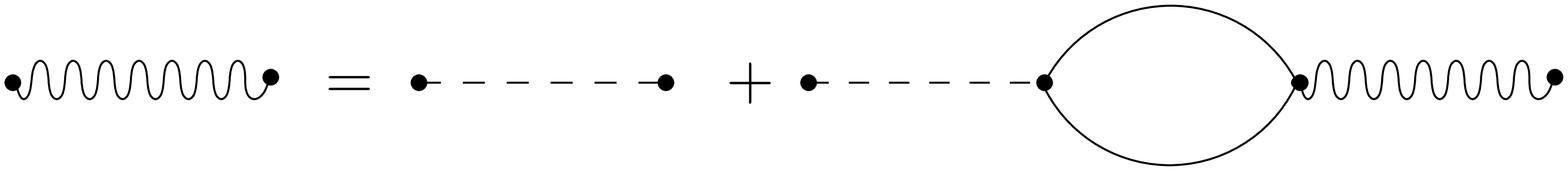}}}
\end{center}
\end{figure}
\noindent
FIG. 1b: Feynman diagrams for $v_{\rm r}=v_0+v_0Qv_{\rm r}$, the screened Coulomb repulsion in RPA.

\begin{figure}[h!]
\begin{center}
\rotatebox{0}{\resizebox{100mm}{!}{\includegraphics{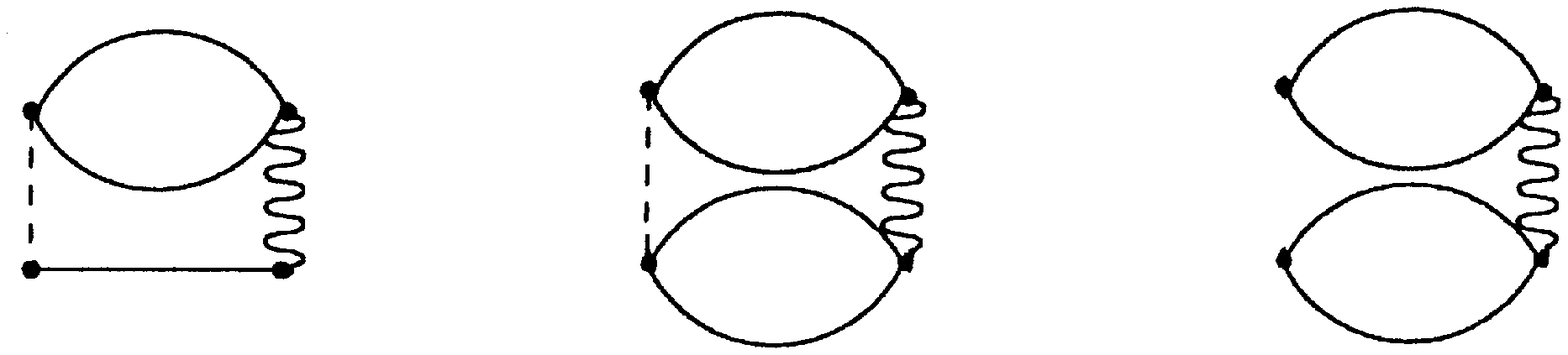}
}}
\end{center}
\end{figure}
\noindent
FIG.1c: Feynman diagrams for the ring-diagram-summed self-energy  
$\Sigma_{\rm r}$ (left) with $n_{\rm r}=G_0\Sigma_{\rm r}G_0$, for the energies $e_{\rm r}$, $v_{\rm r}$ (middle), 
and for $S_{\rm r}$ (right) according to Eqs.  (\ref{c1}), (\ref{c4}), (\ref{c13}), (\ref{c14}). ``Descreening'' $v_{\rm r}\to v_0$
makes the divergent quantities $\Sigma_{\rm 2d}$, $n_{\rm 2d}$, $e_{\rm 2d}$, $v_{\rm 2d}$, $S_{\rm 1d}$. 

\newpage

\begin{figure}[h!]
\begin{center}
\rotatebox{0}{\resizebox{90mm}{!}{\includegraphics{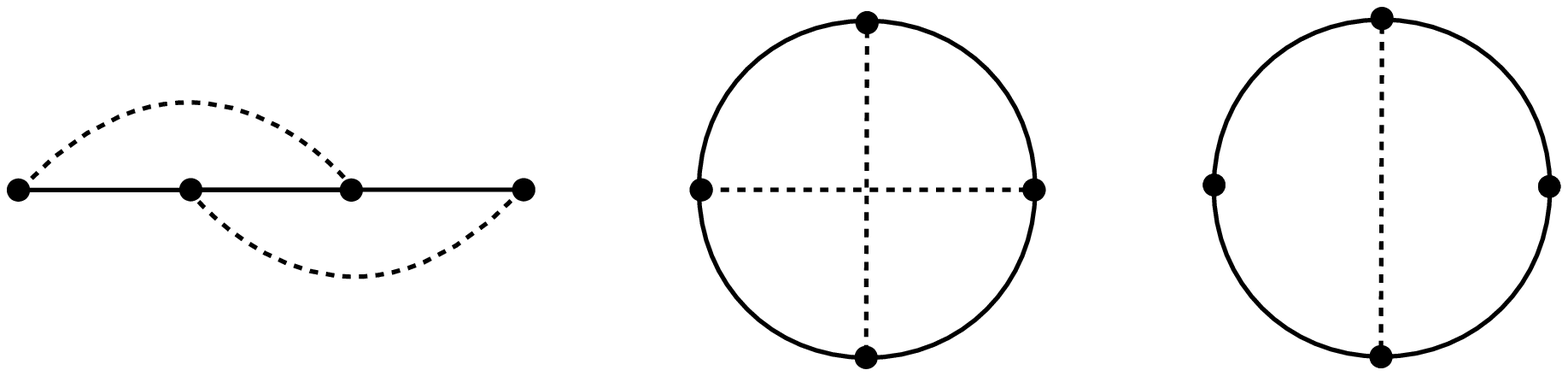}
}}
\end{center}
\end{figure}
\noindent
FIG. 1d: Feynman diagrams for $\Sigma_{\rm 2x}$ (left) with $n_{\rm 2x}=G_0\Sigma_{\rm 2x}G_0$, for the energies 
$e_{\rm 2x}$, $v_{\rm 2x}$ (middle), and for $S_{\rm 1x}$ (right) according to Eqs. (\ref{d1})-(\ref{d5}), (\ref{d7}). 

\begin{figure}[h!]
\begin{center}
{\resizebox{110mm}{!}{\includegraphics{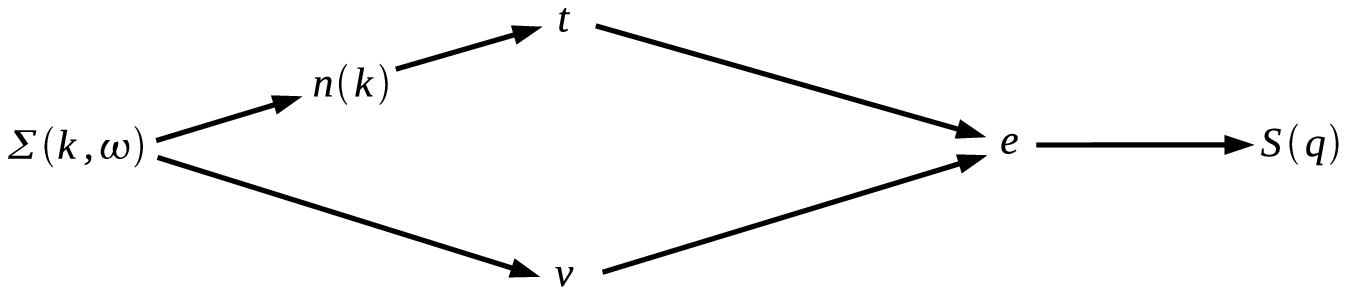}}}
\end{center}
\end{figure}
\noindent
FIG. 2: Terms following from the off-shell self-energy $\Sigma (k,\omega)$ as a functional of $t(\vek k)=\vek k^2/2$ and 
$v(\vek q)=q_c^2/\vek q^2$:
The Migdal formula (\ref{a10}) yields the momentum distribution $n(k)$, from which follows the kinetic energy $t$. The Galitskii-Migdal 
formula (\ref{a11}) yields the potential 
energy $v$. The Hellmann-Feynman functional derivative (\ref{a12}) of the total energy $e=t+v$ yields the SSF $S(q)$, its Fourier 
transform gives the PD $g(r)$. The cases $\Sigma_{\rm x}$, $\Sigma_{\rm r}$, $\Sigma_{\rm 2x}$ are studied in Secs. II, III, IV,
respectively.
 
\begin{figure}[h!]
{\resizebox{90mm}{!}{\includegraphics{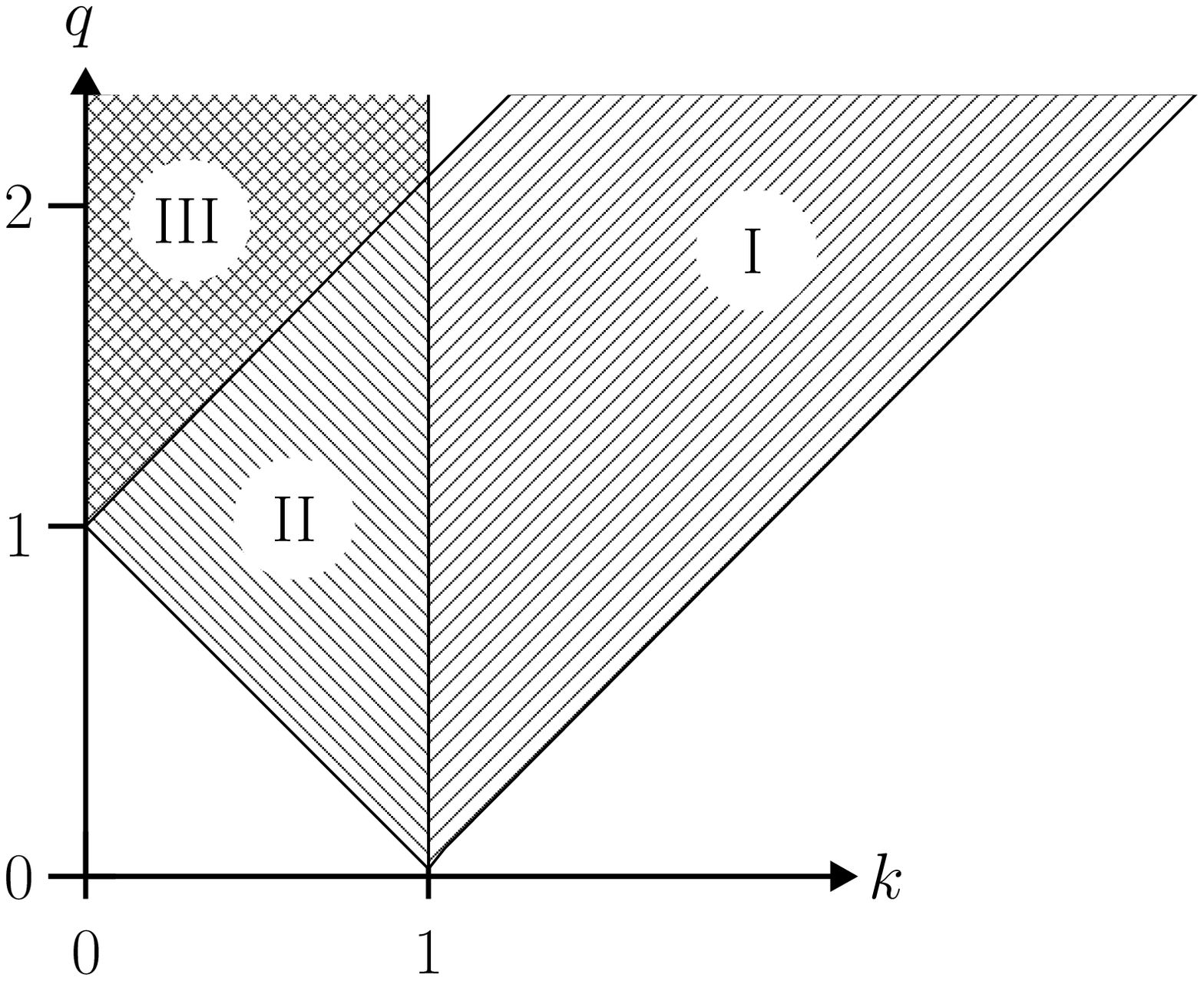}}}
\end{figure}
\noindent
FIG. 3: The momentum distribution (\ref{c38}) and its $k$-$q$-plane.


\begin{thebibliography}{1}
\bibitem{foo}  The Coulomb repulsion takes the form $\alpha r_s/r$, if lengths and energies are measured in units of
$k_{\rm F}^{-1}$ and $k_{\rm F}^2$, respectively. This shows, that $r_s$ is not only a density parameter \cite{Zie1},
but also measures the interaction strength or coupling constant. The Fourier transform of $\alpha r_s/r$ is 
$4\pi\alpha r_s/q^2=\pi^2(q_c^2/q^2)$ with $q$ measured in units of $k_{\rm F}$ and $q_c^2=4\alpha r_s/\pi$ being 
the coupling constant $\alpha r_s$ with a factor $4/\pi$. - Notice that $\tilde e=k_{\rm F}^2e=e/(\alpha r_s)^2$ gives the energy
in a.u., e.g. the energy in zeroth order and the lowest-order exchange energy are ${\tilde e}_0=3/(10\ \alpha^2 r_s^2)$
and ${\tilde e}_{\rm x}=-3/(4 \pi \alpha r_s)$, respectively.
\end{thebibliography}
\end{document}